\documentclass[pra,aps,superscriptaddress,twocolumn,nofootinbib, accepted=2018-01-14]{quantumarticle}
\usepackage{graphicx,color,epstopdf}
\usepackage{amsmath,amsfonts,enumerate,amsthm,amssymb,mathtools}
\usepackage{enumitem}
\usepackage{thmtools,thm-restate}
\usepackage{bbold}
\usepackage{hyperref}
\usepackage{multirow}
\usepackage{subfigure} 
\usepackage{mathdots}  
\usepackage{enumitem}  
\usepackage[utf8]{inputenc}
\usepackage[numbers,sort&compress]{natbib}

\hypersetup{
	colorlinks=true,
	linkcolor=blue,
	citecolor=blue,
	urlcolor=blue
}

\newcommand{\bra}[1]{\ensuremath{\left\langle#1\right|}}
\newcommand{\ket}[1]{\ensuremath{\left|#1\right\rangle}}

\newcommand{\ketbra}[2]{\ensuremath{\left|#1\right\rangle\!\!\left\langle#2\right|}}

\newcommand{\tr}[2]{\mathrm{Tr}_{#1}\left[ #2 \right]}
\newcommand{\iden}{\mathbb{1}}
\renewcommand{\v}[1]{\ensuremath{\boldsymbol #1}}

\newcommand{\be}{\begin{equation}}
\newcommand{\ee}{\end{equation}}

\newcommand{\etho}{\stackrel{\textsc{\tiny ETO}}{\rightarrow}}
\newcommand{\tho}{\stackrel{\textsc{\tiny TO}}{\rightarrow}}
\newcommand{\jc}{\stackrel{\textsc{\tiny JC}}{\rightarrow}}
\newcommand{\notetho}{\stackrel{\textsc{\tiny ETO}}{\nrightarrow}}
\newcommand{\nottho}{\stackrel{\textsc{\tiny TO}}{\nrightarrow}}

\theoremstyle{plain}
\newtheorem{thm}{Theorem}
\newtheorem{corol}[thm]{Corollary}
\newtheorem{lem}[thm]{Lemma}

\theoremstyle{definition}
\newtheorem{defn}{Definition}
\theoremstyle{remark}

\begin{document}
	\title{Elementary Thermal Operations}
	\author{Matteo Lostaglio}
\affiliation{ICFO-Institut de Ciencies Fotoniques, The Barcelona Institute of Science and Technology, Castelldefels (Barcelona), 08860, Spain}  
\author{\'Alvaro M. Alhambra}
\affiliation{Department of Physics and Astronomy, University College London, Gower Street, London WC1E 6BT, UK}
 \author{Christopher Perry}
 \affiliation{QMATH, Department of Mathematical Sciences, University of Copenhagen, Universitetsparken 5, 2100 Copenhagen, Denmark}
 
	\begin{abstract}
To what extent do thermodynamic resource theories capture physically relevant constraints? Inspired by quantum computation, we define a set of elementary thermodynamic gates that only act on 2 energy levels of a system at a time. We show that this theory is well reproduced by a Jaynes-Cummings interaction in rotating wave approximation and draw a connection to standard descriptions of thermalisation. We then prove that elementary thermal operations present tighter constraints on the allowed transformations than thermal operations. Mathematically, this illustrates the failure at finite temperature of fundamental theorems by Birkhoff and Muirhead-Hardy-Littlewood-Polya concerning stochastic maps. Physically, this implies that stronger constraints than those imposed by single-shot quantities can be given if we tailor a thermodynamic resource theory to the relevant experimental scenario. We provide new tools to do so, including necessary and sufficient conditions for a given change of the population to be possible. As an example, we describe the resource theory of the Jaynes-Cummings model. Finally, we initiate an investigation into how our resource theories can be applied to Heat Bath Algorithmic Cooling protocols.
	\end{abstract}
		\maketitle

\section{Introduction}

In recent years, ideas first appearing in the field of physical chemistry \cite{ruch1975diagram, ruch1976principle,mead1977mixing, ruch1978mixing} have been used in conjunction with information theory to analyse thermodynamics in the quantum and nano-scale regimes. Often referred to as the \emph{resource theory} approach to quantum thermodynamics, it has encompassed 2nd law-like statements investigating constraints on the allowed state transformations, e.g. \cite{janzing2000thermodynamic, brandao2011resource, aberg2013truly, horodecki2013fundamental, brandao2013second,skrzypczyk2014work, egloff2015measure, lostaglio2015description,lostaglio2015stochastic,narasimhachar2015low, gemmer2015from, richens2016work},
considerations on the 3rd law \cite{masanes2017general, scharlau2016quantum, wilming2017third} and the construction of fluctuation theorems \cite{halpern2015introducing, aberg2016fully, alhambra2016fluctuating, alhambra2016fluctuating2}. For a more thorough review of results within the field and beyond, see \cite{goold2016role, vinjanampathy2016quantum} and references therein.

Many of these results are based upon the definition of a restricted set of quantum operations known as \emph{thermal operations} (TO). These aim to capture all processes that can be realised \emph{without} an external source of work or coherence and thus derive fundamental limitations and bounds on thermodynamic processes and transformations. More specifically, they consist of all maps on a system with Hamiltonian $H_S$ and in a state $\rho$ that can be written as
\begin{equation}
\label{eq:thermaloperations}
\mathcal{E}(\rho)= \tr{X}{U\left(\rho \otimes \frac{e^{-\beta H_B}}{\tr{}{e^{-\beta H_B}} }\right)U^\dagger}.
\end{equation}
Here $\beta = 1/(kT_B)$ is a fixed inverse temperature of the environment, $H_B$ is an \emph{arbitrary} bath Hamiltonian, $U$ is an energy-preserving unitary satisfying \mbox{$[U,H_S + H_B] =0$}, and typically (but not necessarily) $X=B$, i.e. the partial trace is over the bath degrees of freedom. Note that if \mbox{$[U,H_S + H_B] \neq 0$}, then $U$ requires work to be performed, as can be seen from an argument involving the notion of passivity (see Appendix~\ref{appendix:energyconservation}).

Questions can be raised about the connection of this formalism to relevant experimental scenarios. It is important to stress that giving the definition of thermal operations in the ``Stinespring form'' \cite{nielsen2010quantum}  of Eq.~\eqref{eq:thermaloperations} does \emph{not} imply that an agent needs full control of system-bath interactions and arbitrary baths to implement them; furthermore, there are instances in which a large number of unitaries lead to the same transformation on the system~\cite{brandao2011resource}. Nevertheless, it is not clear what realistic constraints are captured by Eq.~\eqref{eq:thermaloperations}~\cite{halpern2017toward}. In this paper we provide answers to this question by considering subsets of thermal operations which have clear experimental realisations and contrasting them with the full set.

\begin{figure}[t]	
	\includegraphics[width=\columnwidth]{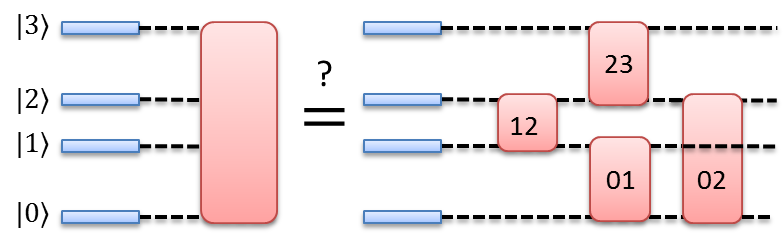}
	\caption{\emph{Decomposing TO}: is it possible to realise any transition allowed by TO by sequences of two-level processes?}
	\label{fig:decomposition}
\end{figure}

In quantum computing, it is common to study which operations can be implemented by sequences of elementary gates. Most famously, every $n$-qubit gate can be implemented using sequences of one and two-qubit gates~\cite{reck1994experimental}. From this perspective it is natural to ask if every transition induced by thermal operations can be implemented through sequences of two-level thermal operations, which we call \emph{elementary thermal operations} (ETOs), see Fig.~\ref{fig:decomposition}.

We analyse the above question by comparing the population dynamics that can be induced by thermal operations and sequences of ETOs. On the one hand, we show that the dynamics induced by an ETO can be satisfactorily captured within a very simple physical model, a Jaynes-Cumming interaction with a single bosonic mode in rotating-wave approximation (RWA); moreover, ETOs are connected to collision models. On the other hand, we show that ETOs constitute a genuinely \emph{new} resource theory -- specifically, one that is more restricted in the allowed transformations and easier to relate to physical models. The presence of a gap between TOs and ETOs already at the classical level may come as a surprise; in fact, it is based on the failure at finite temperature of famous theorems by Muirhead \cite{muirhead1902some}, Hardy, Littlewood and P\'olya \cite{hardy1952inequalities} and Birkhoff \cite{birkhoff1946tres} for doubly-stochastic matrices \cite{marshall2010inequalities}.

Furthermore, we provide necessary and sufficient conditions to compute the population dynamics that are allowed in the theory of ETOs. We then use these results to investigate what further limitations, compared to the thermal operation framework, arise in deterministic work processes. We show that the work of formation under ETOs can be infinite, and we point out a gap between ETOs and TOs in deterministic work extraction. This shows how restrictions in control translate into the impossibility to achieve some of the traditional results of single-shot thermodynamics. In addition, we discuss the role of coherence within the resource theory, showing that ETOs are more limited than thermal operations also with regards to processing of coherence.

The present investigation can be understood as part of a general effort to develop tools to study thermodynamic transformations under restricted control \cite{perry2015sufficient, wilming2016second,scharlau2016quantum,  lekscha2016quantum,mazurek2017preparation}. We show that, while at infinite temperature thermodynamic theories are largely \emph{universal}, i.e. the specific constraints at hand do not matter in terms of the thermodynamic laws arising, this is no longer the case at finite temperature. This implies that every restriction will lead to a different resource theory; luckily, however, some of these can be solved using the same tools we develop here for ETOs. For example, we  can give necessary and sufficient conditions for the population dynamics achievable in the resource theory of 2-level Jaynes-Cummings interactions in RWA, which is a theory strictly contained in ETOs. This is important since, for large enough systems and/or specific choices of physical parameters, we expect the two theories to depart from each other. We also show how to solve a rather general set of thermodynamic resource theories satisfying two assumptions: (a) partial level thermalisation involving 2 levels at a time are (strictly) included and (b) the set of extremal maps in the set of allowed operations is known. Finally, we discuss an application of this framework to an algorithmic cooling protocol, showing that a particular ETO can be used to speed up recently proposed protocols.
  
\section{Elementary thermal operations}

\subsection{Definition and first properties}

A thermal operation will be called \emph{elementary} if it acts non-trivially only on a two-dimensional subspace spanned by two eigenstates of the system Hamiltonian $H_S$. We will denote by $E_i = \hbar \omega_i$ the eigenvalues of $H_S$.

Given a $d$-dimensional system in state $\rho$, denote by $p_i$ the occupation probability of energy level $E_i$, and let \mbox{$\v{p} = (p_0,...,p_{d-1})$}. Note that every thermal operation $\mathcal{E}$ induces a corresponding population dynamics $G$ on $\v{p}$:
\begin{equation}
\label{eq:quantummapstostochasticmatrix}
\mathcal{E}(\rho) = \sigma \Rightarrow G \v{p} = \v{q},
\end{equation}
where $\v{q}$ is the vector of occupation probabilities $q_i$ of $\sigma$ and $G_{k'|k} = \bra{k'}\mathcal{E}(\ketbra{k}{k})\ket{k'}$. $G$ is simply a matrix of transition probabilities $G_{k'|k}$ from energies $\{E_k\}$ to $\{E_{k'}\}$. Since $G_{k'|k} \geq 0$ and $\sum_{k'} G_{k'|k} =1$ (because $\mathcal{E}$ is completely positive and trace-preserving), $G$ is a stochastic matrix. Moreover, one can directly show from Eq.~\eqref{eq:thermaloperations} that $G$ satisfies $G \v{g} = \v{g}$, where $\v{g}$ is the thermal Gibbs distribution of $H_S$ at fixed inverse temperature $\beta$:
\begin{equation}
\label{eq:gibbstate}
\v{g} = \left (\frac{e^{-\beta E_0}}{Z_S}, \dots, \frac{e^{-\beta E_{d-1}}}{Z_S}\right), \quad Z_S= \sum_{i=0}^{d-1} e^{-\beta E_i}.
\end{equation}
Such matrices $G$ are referred to as \emph{Gibbs-stochastic}. In fact, the converse is true: every Gibbs-stochastic matrix can be realised as the population dynamics of (the limit of some sequence of) thermal operations \cite{horodecki2013fundamental, korzekwa2016coherence}. 

Elementary thermal operations give rise to $2$\emph{-level Gibbs-stochastic} matrices $G$. These are a set of matrices exhibiting a particularly simple structure:
\begin{enumerate}
	\item \label{property1} Only two energy levels of the system $(i,j)$ are involved, i.e., $G_{k|k} =1$ for every $k \neq i,j$.
	\item \label{property2} \emph{Detailed balance} holds, i.e. \mbox{$G_{i|j}= e^{-\beta (E_i - E_j)}G_{j|i}$}.
\end{enumerate}
Property~\ref{property1} follows immediately from the definition of ETOs. Property~\ref{property2} follows since $G \v{g} = \v{g}$ is \emph{equivalent} to detailed balance in the case of $2 \times 2$ matrices. 

Note that any population dynamics $G$ induced by an ETO is characterised by the two energy levels $i$ and $j$ on which it acts and a single parameter describing the ``strength'' of the interaction. This can be chosen to be $G_{j|i} \in [0,1]$ if $E_i \geq E_j$. So,
\be
\nonumber
G= \left( \begin{matrix}
 1- G_{j|i}e^{-\beta \hbar \omega_{ij}} & G_{j|i}  \\
 G_{j|i}e^{-\beta\hbar  \omega_{ij}} & 1- G_{j|i}
\end{matrix} \right) \oplus \iden_{\backslash (i,j)},
\ee 
where $\iden_{\backslash (i,j)}$ denotes the identity matrix on every energy level different from $i$, $j$ and $\omega_{ij}:= \omega_i - \omega_j$ is the transition frequency. It is worth stressing that the set of allowed operations we propose does not require the ability to control the energy levels of the system Hamiltonian. We highlight
that changing the energy levels is difficult in practice for many experimental scenarios, but the ETOs do not require this ability. 

Every $2$-level Gibbs stochastic transition matrix can be realised by means of ETOs, without the need for arbitrary bath Hamiltonians:

 \begin{restatable}{lem}{eto}
 \label{lem:eto}
Every population dynamics satisfying properties \ref{property1} and \ref{property2} can be realised by an ETO involving a single-mode bosonic bath.
\end{restatable}

See Appendix~\ref{appendix:elementary} for the simple proof. ETOs hence have a much simpler structure than the generic processes induced by thermal operations and can be implemented using a simple bath. Due to Lemma~\ref{lem:eto}, and since from now on \emph{we will only focus on the population dynamics}, with some abuse of language we will refer to the set of $2$-level Gibbs-stochastic matrices (stochastic matrices satisfying properties \ref{property1} and \ref{property2}) as ETOs. Note that this identification is restricted to the allowed population dynamics. ETOs and Gibbs-preserving maps on two levels do not coincide as sets of quantum channels (see Sec.~\ref{sec:coherence}).

Before we turn to the question of achieving ETOs within simple physical models, let us introduce an ETO of central importance, the $\beta$-swap, and describe how the question posed in Fig.~\ref{fig:decomposition} fits within the general theory of stochastic processes.
  
  \subsection{\texorpdfstring{$\beta$-swaps and partial level thermalisations}{β-swaps and partial level thermalisations}}
\label{sec:betaplt}
  
  A direct swap of population between two energy levels cannot be achieved by thermal operations, since the associated transition matrix is not Gibbs-stochastic (unless $E_i = E_j$ or $\beta = 0$). Nevertheless, there is a transformation that is the thermodynamic analogue of such a swap: the $\beta$\emph{-swap}. It corresponds to an ETO $\beta^{(i,j)}$ among levels $i$ and $j$ ($E_i \geq E_j$) for which \mbox{$G_{j|i} = 1$}:
  \be
  \label{eq:beta-swap}
  \beta^{(i,j)} =  \begin{pmatrix}
  	1- e^{-\beta \hbar \omega_{ij}} & 1 \\
  	e^{-\beta \hbar \omega_{ij}} & 0
  \end{pmatrix} \oplus \iden_{\backslash (i,j)}.
  \ee
An important fact regarding $\beta$-swaps is that if $G$ is any ETO acting on energy levels $i$ and $j$, we have for some $\lambda \in [0,1]$
  	\begin{equation}
  	\label{eq:etobetaswap}
  	G = (1-\lambda) \mathbb{1} + \lambda \beta^{(i,j)}
  	\end{equation}
i.e. every ETO can be written as a convex combination of the identity matrix and a $\beta$-swap. This decomposition will be useful later.

Another decomposition of any ETO can be obtained by combining $\beta$-swaps and \emph{partial level thermalisations} (PLTs). PLTs partially swap the state of two levels $i$, $j$ ($E_i \geq E_j$) for the corresponding thermal probabilities:
\begin{equation}
\label{eq:partialthermalisation}
(q_i,q_j) = (1-\lambda) (p_i,p_j) + \lambda N \v{g}^{(i,j)}, \quad \lambda \in [0,1],
\end{equation}
and $p_k = q_k$ for $k \neq i,j$. Here $N= p_i + p_j$, \mbox{$\v{g}^{(i,j)} = \left(\frac{1}{1+e^{-\beta \hbar \omega_{ij}}},\frac{e^{-\beta \hbar  \omega_{ij}}}{1+e^{-\beta \hbar \omega_{ij}}}\right)$}. PLTs take the form
\begin{equation}
\label{eq:pltmatrix}
T_{\lambda}=\begin{pmatrix}
1-\lambda e^{-\beta \hbar \omega_{ij}} & \lambda  \\
\lambda e^{-\beta \hbar \omega_{ij}}  & 1-\lambda
\end{pmatrix} \oplus \iden_{\backslash (i,j)}.
\end{equation} 
for $\lambda\in [0,1/(1+e^{-\beta\hbar \omega_{ij}})]$. If we take $\lambda$ to be such that \mbox{$\lambda = 1/(1+e^{-\beta\hbar \omega_{ij}})$}, the final state is proportional to a thermal state on the corresponding energy levels and $T_\lambda$ is hence a \emph{full thermalisation} on such levels. Given any ETO $G$, one can directly check that either $G$ is a PLT or there exists a PLT $T_{\lambda}$ and a $\beta$-swap $\beta^{(i,j)}$ such that $G = T_{\lambda} \beta^{(i,j)}$.

\subsection{The infinite temperature limit}  
  
A central question, introduced in Fig.~\ref{fig:decomposition}, is whether every population dynamics induced by thermal operations can be realised by either 
\begin{enumerate}
	\item Sequences of ETOs.
	\item Sequences of ETOs and convex combinations thereof.
\end{enumerate}
A fruitful perspective on these problems comes from the fact that, in the infinite temperature limit $\beta \rightarrow 0$ (or within each energy subspace), thermal operations induce population dynamics $G$ given by doubly-stochastic matrices. These are stochastic matrices $G$ satisfying $G \iden = \iden$. In this limit (studied in Ref.~\cite{gour2015resource}) both of the above questions are answered by two fundamental theorems, one due to Birkhoff \cite{birkhoff1946tres} and the other to Muirhead \cite{muirhead1902some} and Hardy, Littlewood and P\'olya~\cite{hardy1952inequalities}.

Specifically, from Eq.~\eqref{eq:etobetaswap}, one can recognise that ETOs in the infinite temperature limit are convex combinations of the identity and a swap, since $\beta$-swaps are swaps when $\beta \rightarrow 0$. These matrices are simply general $2$-level doubly-stochastic matrices, also known as $T$-\emph{transforms}. They play a crucial role due to the following theorem~\cite{marshall2010inequalities}:
\begin{thm}[Muirhead \cite{muirhead1902some} and Hardy, Littlewood and P\'olya \cite{hardy1952inequalities}]
	\label{thm:muirhead}
	Given $\v{p}$ and $\v{q}$, there exists a doubly-stochastic matrix $G$ such that $G\v{p} = \v{q}$ if and only if there exists a finite sequence of $T$-transforms $G^{(i)}$ such that
	\begin{equation*}
\v{q} = G^{(k)} \cdots G^{(1)} \v{p}.
	\end{equation*}
\end{thm}
At infinite temperature, this theorem answers the first question above, since it shows that all transformations induced by thermal operations can also be realised by finite sequences of ETOs.
 
 If we also allow convex combinations, we can take advantage of a fundamental theorem by Birkhoff to strengthen the above result:
 \begin{thm}[Birkhoff \cite{birkhoff1946tres}]
 	\label{thm:birkoff}
 	Every doubly-stochastic matrix can be written as a convex combination of permutations.
 \end{thm}
 Since permutations are sequences of swaps,  Birkhoff's theorem implies that, if we allow convex combinations, then not only do ETOs induce the same transformations $\v{p} \rightarrow \v{q}$ as thermal operations at infinite temperature, but they can in fact simulate every transition matrix generated by a thermal operation, i.e. every doubly-stochastic matrix.\footnote{This is strictly stronger. In fact, the set of matrices that can be realised through sequences of $T$-transforms is strictly smaller than the set of all doubly-stochastic matrices, see \cite[Chapter~2]{marshall2010inequalities}.}
 
\subsection{\texorpdfstring{Failure of finite temperature extensions: $\textrm{ETO} \neq \textrm{TO}$}{Failure of finite temperature extensions: ETO≠TO}}

 \label{sec:failure}
 We can now see that asking if ETOs coincide with thermal operations (in terms of the achievable transformations or in terms of the set of transition matrices) corresponds to asking: do the two central theorems \ref{thm:muirhead} and \ref{thm:birkoff} in the theory of doubly stochastic matrices extend from infinite to finite temperatures? We will show that, in both the weaker and stronger sense,
 \begin{equation*}
 \textrm{elementary thermal operations} \neq \textrm{thermal operations},
 \end{equation*}
 and hence both Theorems \ref{thm:muirhead} and \ref{thm:birkoff} fail at finite temperatures, i.e. for Gibbs-stochastic matrices. The consequence is that ETOs define a \emph{genuinely new resource theory of thermodynamics}. We leave the proof of these facts to Section \ref{sec:ETO theory} of this manuscript.
 
  Knowing that ETOs provide a new resource theory based on a simple set of allowed operations, two natural questions arise:
  \begin{enumerate}
  	\item Can we realise every ETO within simple physical models?
  	\item What transformations can be realised by combining ETOs?
  \end{enumerate}  
  We begin by tackling the first question and show that, in contrast to thermal operations, ETOs are easily related to simple physical models.

\section{Physical models for elementary thermal operations}
\label{sec:etophysicalmodels}

While ETOs only require interactions with a single mode bosonic bath (Lemma~\ref{lem:eto}), we are interested here in considering  physically realistic models. This may require additional restrictions, e.g. on the dimension of the bath or on the allowed system-bath interactions. The results of Ref.~\cite{scharlau2016quantum} imply that not all ETOs can be achieved with a finite-size bath. Here we consider instead some natural limitations on the allowed system-bath interactions. We begin this study by analysing if the Jaynes-Cummings (JC) model reproduces the set of ETOs to a satisfactory extent. For many practical purposes, which is what interests us in providing a physical model, we shall see that the answer is affirmative.

\subsection{Jaynes-Cummings model} 
\label{sec:JC}

One of the core descriptions of the interaction between matter and radiation is the Jaynes-Cummings (JC) model \cite{jaynes1963comparison, shore1993jaynes}. Hence, we wish to know if it can provide a natural model for ETOs. 

For simplicity, consider a system with Hamiltonian $H_S = \sum_{k=0}^{d-1} E_k \ketbra{E_k}{E_k}$ whose relevant transition frequencies are all well separated from each other and non-degenerate, i.e.
\begin{equation}
\label{eq:nondegenerate}
E_{j} - E_{i} \neq E_{j'} - E_{i'} \quad \textrm{if} \quad  (j,i) \neq (j',i').
\end{equation}
Assume one is allowed to couple every transition $\omega_{ij}$ of the system to a single-mode bosonic bath with Hamiltonian \mbox{$H_B = \sum_{n=0}^\infty n \hbar \omega_{ij} \ketbra{n}{n}$}. The interaction is given by a resonant JC Hamiltonian $H_{\textrm{JC}}$ in rotating wave approximation, 
\begin{equation*}
H_{\textrm{JC}} = g( \sigma_+ \otimes a + \sigma_- \otimes a^\dag).
\end{equation*}
Here $a^\dag$, $a$ are creation/annihilation operators on the bath and $\sigma_{+} = \ketbra{E_j}{E_i}$, $\sigma_- = \sigma_+^\dagger$ excite/de-excite the relevant transition. 

Without loss of generality let us denote the two levels of the system involved by $\ket{E_0}$ and $\ket{E_1}$ and set \mbox{$\hbar \omega := E_1 - E_0 >0$}. One can compute (\cite{aberg2014catalytic}, Appendix~E.2)
 \begin{eqnarray*}
 	e^{-\frac{i t}{\hbar} H_{\textrm{JC}}} &=& \sum_{n=1}^\infty \sum_{k,k'=0}^1 u^{(n)}_{k,k'}(s) \ketbra{E_k}{E_{k'}} \otimes \ketbra{n-k}{n-k'} \\
 	&+& \ketbra{E_0}{E_0} \otimes \ketbra{0}{0},
 \end{eqnarray*}
 where $s=gt/\hbar$ and, for each $n$, $u^{(n)}_{k,k'}(s)$ are matrix elements of the unitary
 \be
 \nonumber
 U^{(n)}(s) = \left[ \begin{matrix}
 	\cos (s \sqrt{n}) & -i \sin (s\sqrt{n}) \\
 	-i \sin(s \sqrt{n}) & \cos (s\sqrt{n})
 \end{matrix} \right].
 \ee 
 Take the system and bath to be in the initial state \mbox{$\rho \otimes e^{-\beta H_B}/Z_B$}, where $Z_B = (1-e^{-\beta \hbar \omega})^{-1}$. One then has that the dynamics of the system in the interaction picture is described by\footnote{To realise any given $s$, one should take $g$ small enough and $t$ large enough to avoid the short times/strong couplings regime where the model breaks down.}
 \be
 \nonumber
 \rho \mapsto \tr{B}{e^{-\frac{i t}{\hbar} H_{\textrm{JC}}} \left(\rho \otimes \frac{e^{-\beta H_B}}{Z_B}\right) e^{\frac{i t}{\hbar} H_{\textrm{JC}}}}.
 \ee
 Since $[H_{\textrm{JC}}, H_S + H_B] =0$, this is an ETO. The transformation induced on the vector $\v{p}$ of occupation probabilities is a stochastic process $G$ whose transition probabilities are:
 {\small \be
 	\nonumber
 	G_{1|0}(s) = \sum_{n=1}^\infty \sin^2(s \sqrt{n}) \frac{e^{-\beta \hbar \omega n}}{Z_B}, \quad G_{0|0} = 1- G_{1|0},
 	\ee
 	\be
 	\label{eq:JCtransitions}
 	G_{0|1}(s) = \sum_{n=1}^\infty \sin^2(s \sqrt{n}) \frac{e^{-\beta \hbar \omega (n-1)}}{Z_B}, \quad G_{1|1} = 1- G_{0|1}.
 	\ee
 }
Of course, $G_{k|k} =1$ for every $k \neq 0,1$. As described before, $G$ can be parametrised by $G_{0|1} \in [0,1]$. The question is if this parameter interval can be covered in the JC model.  All values between $0$ and the maximum achievable $G_{0|1}(s)$ (denoted by $G_{{\rm max}}(\beta \hbar \omega)$) can be achieved, since  $G_{0|1}(s)$ is a continuous function in $s$ and it takes the value $0$ at $s=0$.
 
 First, let us investigate the low temperature limit. For every $s$ we have \mbox{$G_{0|1}(s) \geq (1- e^{-\beta \hbar \omega})\sin^2(s)$}. Hence, in the zero temperature limit \mbox{$\beta \rightarrow \infty$}, $G_{{\rm max}}(\beta \hbar \omega) \rightarrow 1$. In this limit every ETO can be reproduced within the JC model with arbitrary accuracy.

\begin{figure}[t]
	\includegraphics[width=\columnwidth]{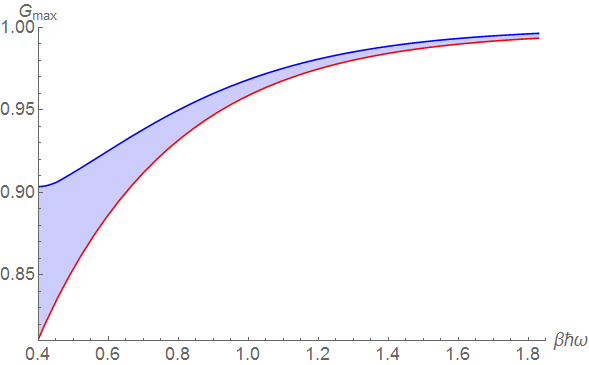}
	\caption{\emph{Realising ETOs in the simplest Jaynes-Cummings model:} to achieve every ETO one needs to be able to obtain all de-exciting probabilities $G_{j|i} \in [0,1]$, $ \omega = \omega_j - \omega_i > 0$. The JC models realises every transition probability within $[0,G_{{\rm max}}(\beta \hbar \omega)]$, where $G_{{\rm max}}(\beta \hbar \omega)$ lies somewhere within the shaded area. Hence, far from the high temperature limit, almost all ETOs can be achieved within the JC model. Here the blue line corresponds to the upper bound given in Eq.~\eqref{eq:upperbound}, while the red line is the lower bound is obtained by truncating the infinite series in Eq.~\eqref{eq:JCtransitions}.}
	\label{fig:achievable_region}
\end{figure}

Now consider any finite temperature. Setting $\bar{\beta}= \beta \hbar \omega$ one can prove (see Appendix~\ref{appendix:bound})
\begin{small}
\begin{equation}\label{eq:upperbound}
G_{{\rm max}}(\bar{\beta}) \leq \left\{
\begin{array}{ll}
\frac{1}{16} \left(8 e^{-\bar{\beta}}-e^{2 \bar{\beta}}+e^{3 \bar{\beta}}+8\right), \; \textrm{for } \bar{\beta} \in [0, \frac{\log(4)}{3}], \\
e^{-4 \bar{\beta}}-e^{-3 \bar{\beta}}+1, \quad \quad \quad \quad \quad \textrm{for } \bar{\beta} \geq \frac{\log(4)}{3}.
\end{array}
\right.
\end{equation}
\end{small}
On the other hand, a rough lower bound on $G_{{\rm max}}(\bar{\beta})$ (good when temperatures are not too high) can be obtained by truncating the sum in Eq.~\eqref{eq:JCtransitions} at some finite $m_{{\rm max}}$, and setting every other term to zero. We take $m_{{\rm max}}=12$. Then numerics suggest to choose $s=98.92$. This, together with the previous upper bounds, provides a region where $G_{{\rm max}}(\bar{\beta})$ must lie as a function of $\bar{\beta}$, see Fig.~\ref{fig:achievable_region}.

 The finite temperature scaling is rather favourable: at room temperature $T_B=300K$ and frequencies $\omega/(2\pi)$ larger than \mbox{$10^{13}$ Hz}, and at millikelvin temperatures with frequencies larger than \mbox{$10^8$ Hz}, one has \mbox{$G_{{\rm max}}(\beta \hbar \omega)>0.989$}. Hence Fig.~\ref{fig:achievable_region} shows that, even though not every ETO can be realised within the JC in RWA, a vast majority of them can be achieved, especially at lower temperatures. This establishes the connection at the level of single operations,
 \begin{equation*}
 \textrm{ETOs} \leftrightarrow \textrm{Jaynes-Cummings model}.
 \end{equation*}

Of course, to implement a sequence of ETOs one still requires considerable frequency control as we must be able to separately couple to every transition frequency of the system. This is feasible to the extent in which we are interested in the thermodynamics of small enough systems. In Section~\ref{ap:HBAC}, we give some suggestions relating to protocols exploiting this set of operations.

\subsection{Collision models}
\label{sec:collision}

We now turn to investigating the connection between the framework of ETOs and thermalisation models, widely used in quantum thermodynamics \cite{kosloff2013quantum}.

Assume again for simplicity that Eq.~\eqref{eq:nondegenerate} holds. Consider a \emph{collision model} of thermalisation of a system in the presence of a large thermal bath \cite{ziman2005description}. If we describe the bath as a collection of qubits with various energy gaps, in this simple model a thermal particle from the bath approaches, interacts with a two-level subspace of the system and scatters away; the interaction conserves energy and the particle is lost in the bath~\cite{scarani2002thermalizing}. Each ``collision'' is then an ETO. Reasoning as in Ref.~\cite{scarani2002thermalizing}, one finds that the combined effect of many identical weak interactions on each two-level subsystem, initially  described by a population $\v{p}(0)$, leads to an exponential relaxation to the thermal distribution $\v{g} $~ (Appendix~\ref{appendix:collision}):
\begin{equation}
\label{eq:partiallevelthermalisation}
\v{p}(t) = e^{-t/\xi} \v{p}(0) + N (1-e^{-t/\xi})\v{g},
\end{equation}
where $\xi \geq 0$ and $N$ is the normalisation of $\v{p}(0)$. This relates ETOs to the collision model.

Alternatively, from the master equation point of view~\cite{breuer2002open}, we can consider Davies quantum dynamical semigroups, which are routinely used as a simple model of a system weakly interacting with a large thermal bath~\cite{davies1974markovian}, or in the low-density limit \cite{dumcke1985low}. On $2$-level systems, a full characterisation of their action exists~\cite{roga2010davies} and it coincides with Eq.~\eqref{eq:partiallevelthermalisation}. Hence, the collision model of ETOs reproduces Davies maps on qubits.

 Comparing Eq.~\eqref{eq:partiallevelthermalisation} with Eq.~\eqref{eq:partialthermalisation}, we see that the models described effectively realise a subclass of ETO dynamics, i.e. partial level thermalisations (defined in Sec.~\ref{sec:betaplt}). These can be understood as \emph{Markovian ETOs}, defined as the subset of ETOs whose induced transition matrices $G$ are embeddable stochastic matrices, i.e. there is some generator $L$ and $t \geq 0$ such that $G = e^{L t}$ \cite{davies2010embeddable}. For $G$ to be a valid stochastic matrix, $L$ must satisfy $L_{i|j} \geq 0$ for $i \neq j$ and $\sum_{i} L_{i|j} = 0$ for all $j$  \cite{davies2007linear}. Detailed balance is equivalent to \mbox{$L_{i|j} = e^{-\beta \hbar (\omega_i - \omega_j)}L_{j|i}$}, where $(i,j)$ are the two levels involved. Computing $e^{Lt}\v{p}$ then gives Eq.~\eqref{eq:partiallevelthermalisation}. 
 
\begin{figure}[t]
	\centering
	\includegraphics[width=0.7\columnwidth]{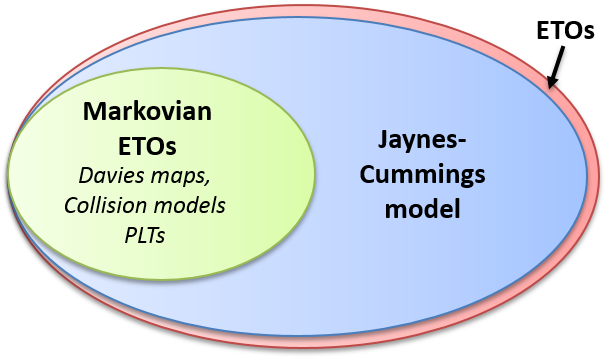}
	\caption{\emph{ETOs and physical models:} The set of ETOs is closely reproduced by the Jaynes-Cummings model for a wide set of physically relevant parameters and with growing precision as temperatures lowers. The Markovian subset of ETOs, on the other hand, is connected to simple thermalisation models. }
	\label{fig:ETO_models}
\end{figure}

 In conclusion, Markovian ETOs can be related to two different physical models: collision models and Davies maps. For a summary of the relations between ETOs and physical models, see Fig.~\ref{fig:ETO_models}.

\section{The resource theory of elementary thermal operations} \label{sec:ETO theory}

We now describe in more detail the resource theory of ETOs and provide necessary and sufficient conditions for a transformation to be possible under this set. In particular, we will present explicit results for qutrit systems, since these can be readily visualised, and discuss a typical thermodynamic task, work extraction. Let us begin with
\begin{defn}[Resource theory of elementary thermal operations]
	The resource theory of ETOs allows all transformations that can realised by sequential application of ETOs and convex combinations thereof. When such operations exist transforming $\v{p}$ into $\v{q}$, we denote this by $\v{p} \etho \v{q}$.
\end{defn}
Allowing convex combination simply corresponds to the ability to condition the choice of our protocol on some random variable (like a coin toss). It is difficult to envisage an experimental setup in which this should be forbidden. If no sequence of ETOs (no matter how long) nor convex combinations thereof can transform $\v{p}$ into $\v{q}$, we will write \mbox{$\v{p} \notetho \v{q}$}. We will use a similar notation if a transformation is possible by means of thermal operations: \mbox{$\v{p}\tho \v{q}$} (for any other set $A$, \mbox{$\v{p} \stackrel{\textsc{\tiny A}}{\rightarrow} \v{q}$}).

\subsection{Counterexample: ETOs do not have the same power as thermal operations}
\label{sec:counterexample}

As anticipated in Sec.~\ref{sec:failure}, the resource theory of ETOs is distinct from thermal operations, in that we are able to show that there are $\v{p}$ and $\v{q}$ such that $\v{p} \tho \v{q}$ but $\v{p} \notetho \v{q}$.\footnote{A result that can be proven to be equivalent to the one we present in this subsection was claimed in Ref.~\cite{veinott1971least}. However, to the best of our knowledge, its proof was never published \cite[Chapter 14B]{marshall2010inequalities}.}

To do this, let us recall the definition of a \emph{monotone} in a resource theory: it is a quantity that never decreases (or never increases) under the allowed operations. We shall show that there is a monotone in the resource theory of ETO that is not a monotone for thermal operations, namely the size of the support of~$\v{p}$. 

\begin{figure}[t] 
\includegraphics[width=0.9\columnwidth]{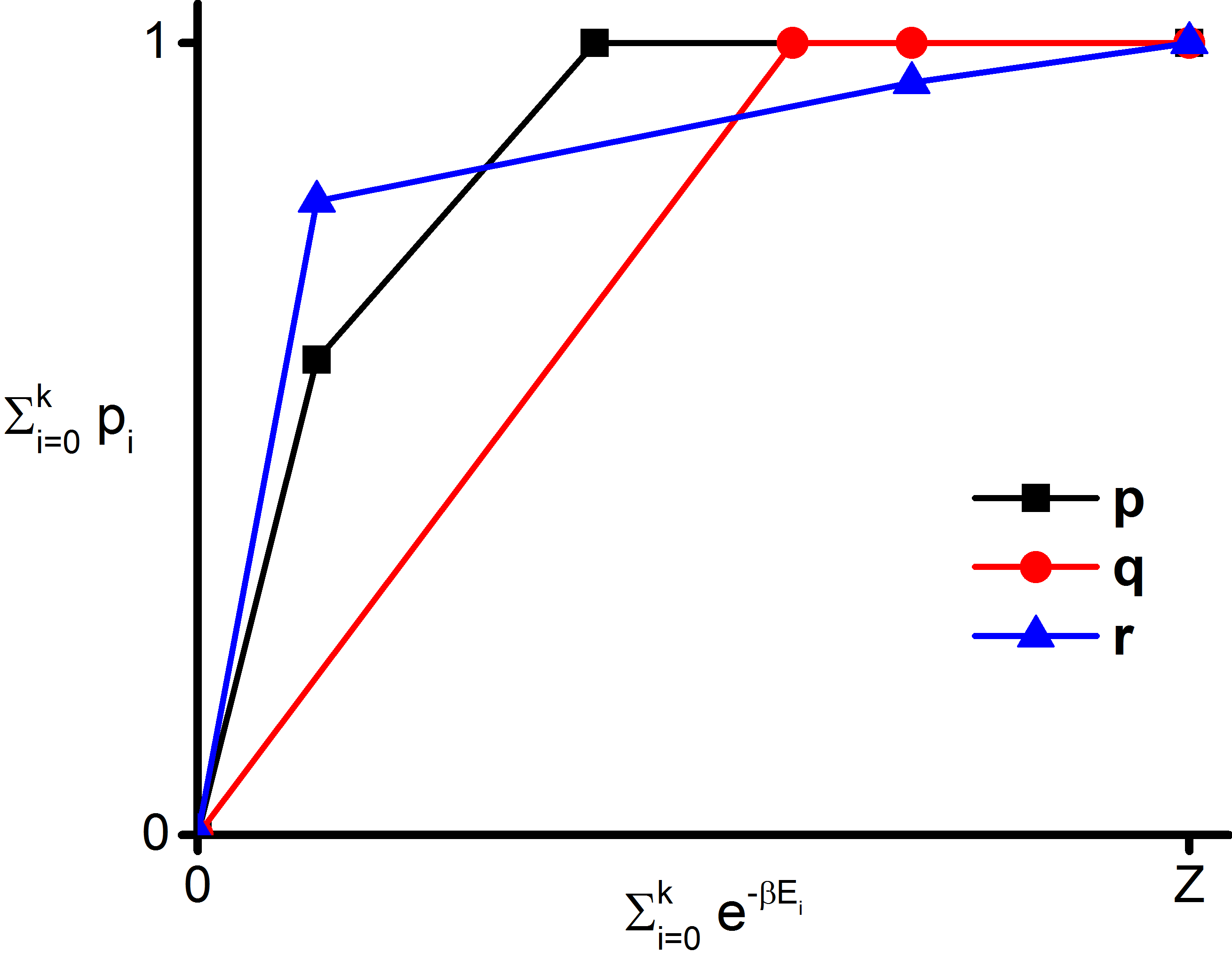}
\caption{\emph{Thermo-majorisation curves:} We present the curves for three probability distributions \mbox{$\v{p} = (0,0.4,0.6)$}, \mbox{$\v{q}= (1,0,0)$} and \mbox{$\v{r} = (0.15, 0.05, 0.8)$}, each associated with the same Hamiltonian $H$ satisfying $e^{-\beta E_0}=1.5$, $e^{-\beta E_1}=0.7$ and $e^{-\beta E_2}=0.3$. Here, $\v{p} \tho \v{q}$, as the thermo-majorisation curve of $\v{p}$ is never below that of $\v{q}$. However, since the thermo-majorisation curve of $\v{r}$ crosses those of $\v{p}$ and $\v{q}$, we have $\v{r} \nottho \v{p}, \v{q}$ and $\v{p} , \v{q} \nottho \v{r}$. The states $\v{p}$ and $\v{q}$ form the basis of Corollary~\ref{corol:counter}. Note that the size of the support of $\v{p}$ is $2$, while for $\v{q}$ it is $1$.}
\label{fig:thermo_maj}
\end{figure}

The second concept we need to introduce to build the counterexample is an important tool in the study of thermal operations: \emph{thermo-majorisation}. Given an initial state $\v{p}$ with corresponding energy levels \mbox{$\v{E} = (E_0,...,E_{d-1})$}, the associated \emph{thermo-majorisation curve} is constructed as follows:
\begin{enumerate}
\item Reorder $\v{p}$ and $\v{E}$ so that $p_i e^{\beta E_i}$ is in non-increasing order. Such an ordering is called the \emph{$\beta$-order} of~$\v{p}$.
\item Plot the ordered points \mbox{$\left\{\sum_{i=0}^{k}e^{-\beta E_i},\sum_{i=0}^{k} p_i\right\}_{k=0}^{d-1}$} together with the point $\left(0,0\right)$ and connect them piecewise linearly to form a concave curve - the thermo-majorisation curve of $\v{p}$.
\end{enumerate}
Given two probability distributions $\v{p}$ and $\v{q}$ associated with the same energy levels, we say that $\v{p}$ thermo-majorizes $\v{q}$ if the thermo-majorisation curve of $\v{p}$ is never below that of $\v{q}$ (examples are given in Fig.~\ref{fig:thermo_maj}). Recalling that a Gibbs-stochastic matrix $G$ with $G \v{p} = \v{q}$ exists if and only if $\v{p}$ thermo-majorises $\v{q}$ \cite{ruch1980generalisation}, and the fact that every such $G$ can be realised through TOs, it follows \cite{horodecki2013fundamental}
\begin{equation}
\label{eq:thermomajorisation}
\v{p} \tho \v{q} \quad \Leftrightarrow \quad \v{p} \; \; \textrm{thermo-majorises}  \; \; \v{q}.
\end{equation} 
This implies that under TOs the height of the thermo-majorisation curve at each value on the $x$-axis cannot increase; furthermore, the decrease of these heights is not only necessary but also \emph{sufficient} for $\v{p} \tho \v{q}$.

As ETOs form a subset of thermal operations, the heights of a thermo-majorisation curve are also monotones in the theory of ETOs. However, we shall now show that additional monotones exist:
\begin{lem}\label{lem:supp}
Given a probability distribution $\v{p}$, let  $|\textrm{supp}\left(\v{p}\right)|= \left\{\textrm{number of } i:p_i>0\right\}$. Then $\v{p} \etho \v{q}$ implies
\begin{equation}
|\textrm{supp}\left(\v{p}\right)|\leq |\textrm{supp}\left(\v{q}\right)|,
\end{equation}
i.e. the size of the support is a monotone in the theory of ETOs.
\end{lem}
\begin{proof}
First note that given two probability distributions $\v{p}_1$ and $\v{p}_2$ and $\lambda\in\left[0,1\right]$:
{\small
\begin{equation*}
|\,\textrm{supp}\left(\lambda\v{p}_1+\left(1-\lambda\right)\v{p}_2\right)|\geq \max\left\{|\,\textrm{supp}\left(\v{p}_1\right)|,|\,\textrm{supp}\left(\v{p}_2\right)|\right\},
\end{equation*}}so taking convex combinations of ETOs can only increase the size of the support. Hence, we can assume $\v{q} = M \v{p}$, with $M$ a sequence of ETOs. Furthermore, when combined with the decomposition of an ETO given in Eq.~\eqref{eq:etobetaswap}, this implies that $M$ can be written as a convex combination of sequences of $\beta$-swaps. Then, $\v{q}$ will have support larger or equal to the support of the distributions obtained by applying such sequences to $\v{p}$. We thus need to consider only the effect of $\beta$-swaps $\beta^{(i,j)}$. As the matrix associated with $\beta$-swaps, given in Eq.~\eqref{eq:beta-swap}, has full rank, it cannot decrease the size of the support.
\end{proof}
These considerations lead us to the following
\begin{corol} \label{corol:counter}
There exists $\v{p}$ and $\v{q}$ such that $\v{p} \tho \v{q}$ but $\v{p}\notetho \v{q}$.
\end{corol}
\begin{proof}
Consider the three-level Hamiltonian with energy levels such that $e^{-\beta E_1}+e^{-\beta E_2}\leq e^{-\beta E_0}$ together with the initial state $\v{p}$ such that $p_0=0$ and the final state $\v{q}$ such that $q_0=1$. For example, \mbox{$\v{p} = (0,0.4,0.6)$}, \mbox{$\v{q}= (1,0,0)$} and $e^{-\beta E_0}=1.5$, $e^{-\beta E_1}=0.7$, $e^{-\beta E_2}=0.3$, as presented in Fig.~\ref{fig:thermo_maj}. As the support of $\v{q}$ in smaller than $\v{p}$, $\v{p} \notetho \v{q}$. However, as shown in Fig.~\ref{fig:thermo_maj}, $\v{p}$ thermo-majorizes $\v{q}$ and hence $\v{p} \tho \v{q}$.
\end{proof}

To make the counterexample robust to the unavoidable experimental imprecisions, one would like a stronger statement: there are transformations allowed by thermal operations that cannot be approximated arbitrarily well in the resource theory of ETOs. Specifically, we want to exclude that there exists a sequence of distributions $\v{q}_\epsilon$ with $\v{q}_\epsilon$ $\epsilon$-close to $\v{q}$ and such that $\v{p} \etho \v{q}_\epsilon$ for every $\epsilon > 0$. This stronger result follows from the fact that set of $\v{q}$ that can be achieved in the resource theory of ETOs is a closed set; in turn, this is a by-product of the considerations of the next section, in which we develop necessary and sufficient conditions for a transformation to be possible in the resource theory of ETOs.

\subsection{Necessary and sufficient conditions} \label{sec:nec&suff}

Given some initial state $\v{p}$, we will now focus on finding all possible states that can be reached from it in the theory of ETOs. In other words, we want to study

	\begin{defn}[ETO cone]
		Given an initial state $\v{p}$, let $\mathcal{C}_{\rm ETO}(\v{p})$ be the set of all $\v{q}$ such that $\v{p} \etho \v{q}$. 
	\end{defn}
	
The strategy will be to characterise the extremal points of the above convex set. We can show that the ETO cone can be completely explored by convex combinations of a finite number of $\beta$-swaps $\beta^{(i,j)}$:
\begin{thm}[Extremal points of ETO cone]
	\label{thm:extremal}
	All extremal points $\v{q}$ of $\mathcal{C}_{\rm ETO}(\v{p})$ can be written as
	\begin{equation}
	\label{eq:betasequence}
\v{q} = \beta^{(i_n,j_n)} \cdots \beta^{(i_1,j_1)} \v{p},
	\end{equation}
	where $n \leq d!$. 
\end{thm}
\begin{proof}
	Eq.~\eqref{eq:etobetaswap} allows us to distinguish the identity and $\beta$-swaps as \emph{extremal} ETOs. The first step of the proof will be to show that if $\v{q}$ is extremal, then it can be obtained through some sequence of $\beta$-swaps, as in Eq.~\eqref{eq:betasequence}. By assumption, since \mbox{$\v{q} \in \mathcal{C}_{\rm ETO}(\v{p})$}, we can write \mbox{$\v{q} = \sum_i a_i M_i \v{p}$}, where each $M_i$ is a sequence of ETOs and $\{a_i\}$ is a probability distribution. Each sequence of ETOs $M_i$ can contain both extremal and non-extremal ETOs. Any non-extremal ETO can be decomposed using Eq.~\eqref{eq:etobetaswap}, obtaining
	\begin{equation}
	\label{eq:ccsequencesbetaswaps}
	\v{q} = \sum_{x} b_x \beta^x\v{p},
	\end{equation}
	where each $\beta^x$ denotes a sequence of $\beta$-swaps and $\{b_x\}$ is some probability distribution, with $b_x \neq 0$ for all $x$. We must then have $\beta^x\v{p} = \v{q}$ for every $x$, otherwise $\v{q}$ is not extremal. This completes the first part of the proof. 
	
	The second part of the proof derives the bound \mbox{$n \leq d!$} on the length $n$ of the required sequence of $\beta$-swaps. By contradiction, assume that the length of the shortest sequence of $\beta$-swaps $\beta^{\rm min}$ such that $\beta^{\rm min} \v{p} = \v{q}$ is $r > d!$. For \mbox{$k=1,...,r$}, define \mbox{$\v{q}^k = \beta^{(i_k,j_k)} \cdots \beta^{(i_1,j_1)} \v{p}$} to be the intermediate states obtained by applying the elements of the sequence $\beta^{\rm min}$. Without loss of generality, we can assume that $\v{q}^k \neq \v{q}^{k'}$ for every $k$, $k'$, otherwise there would be a sequence of length $r' < r$ mapping $\v{p}$ into $\v{q}$, in contradiction with our hypothesis. 
	
	For any $\v{q}^k$, define the $\beta$-ordering of $\v{q}^k$ as a vector of indexes $\v{\pi}^k$, obtained by a permutation of \mbox{$(0,...,d-1)$}. Since there are only $d!$ distinct $\beta$-orderings, after $n> d!$ $\beta$-swaps we necessarily have $k_1$, $k_2$ with $k_1 < k_2$ and 
	\begin{enumerate}
		\item $\v{q}^{k_1} \neq \v{q}^{k_2}$ and $\v{\pi}^{k_1} = \v{\pi}^{k_2}$,
		\item  $\v{q}^{k_2}$ can be obtained from $\v{q}^{k_1}$ through a thermal operation.
	\end{enumerate}
	Note that the second property holds trivially, since sequences of ETOs are also sequences of TOs and the latter are closed under composition. From the two properties above, applying Theorem~12 of Ref.~\cite{perry2015sufficient}, we deduce that there exists a sequence of PLTs transforming $\v{q}^{k_1}$ into $\v{q}^{k_2}$. As in the first part of the proof, since PLTs are non-extremal, we can decompose them through $\beta$-swaps using Eq.~\eqref{eq:etobetaswap}. Hence, $\v{q}^{k_2} = \sum_{y} c_y \beta^y\v{q}^{k_1}$, where as before $\beta^y$ denotes a sequence of $\beta$-swaps. This allows us to substitute part of the sequence $\beta^{\rm min}$, obtaining an expression with the following structure:
	\begin{equation*}
	\v{q} = \beta^{\rm min } \v{p} = \beta^{x_2} \left(\sum_y c_y \beta^{y}\right) \beta^{x_1} \v{p},
	\end{equation*}
	where $\beta^{x_1}\v{p} = \v{q}^{k_1}$, $\beta^{x_2}\v{q}^{k_2} = \v{q}$ and $\{c_y\}$ is some probability distribution with $c_y \neq 0$ for all $y$. Note that the sum of the lengths of $\beta^{x_1}$ and $\beta^{x_2}$ is at most $r-1$. Since $\v{q}$ is extremal we must have, for every~$y$,
	\begin{equation*}
\v{q} = \beta^{x_2}  \beta^{y} \beta^{x_1} \v{p}.
	\end{equation*}
	Note however that there is a $y$ such that $\beta^{y}$ is the identity, i.e. a sequence of length zero. It follows that there is a sequence of $\beta$-swaps mapping $\v{p}$ to $\v{q}$ and of length at most $r-1$. This is in contradiction with our hypothesis. Hence, it must be that $n \leq d!$.
\end{proof}

The above theorem gives, in particular, necessary and sufficient conditions for determining whether \mbox{$\v{p} \etho \v{q}$}. One can proceed as follows:
\begin{enumerate}
	\item Construct the extremal points of $\mathcal{C}_{\rm ETO}(\v{p})$ using Theorem~\ref{thm:extremal}.
	\item $\v{p} \etho \v{q}$ if and only if $\v{q} \in \mathcal{C}_{\rm ETO}(\v{p})$. Determining the latter, given that the extremal points of $\mathcal{C}_{\rm ETO}(\v{p})$ are known, can be solved by a linear program.
\end{enumerate}

We illustrate the above general results by comparing the ETO cone with the TO cone\footnote{For completeness, we show how to construct the TO cone in Appendix~\ref{appendix:TO Cone}} for $d=3$, see Fig.~\ref{fig:ETO+TO_Cone}.

We also note that Theorem~\ref{thm:extremal} implies that $\mathcal{C}_{\rm ETO}(\v{p})$ is a simplex with a finite number of vertices. As such, it is a closed set (as noted at the end of Section~\ref{sec:counterexample}). 
 
\begin{figure}[t] 
\includegraphics[width=1\columnwidth]{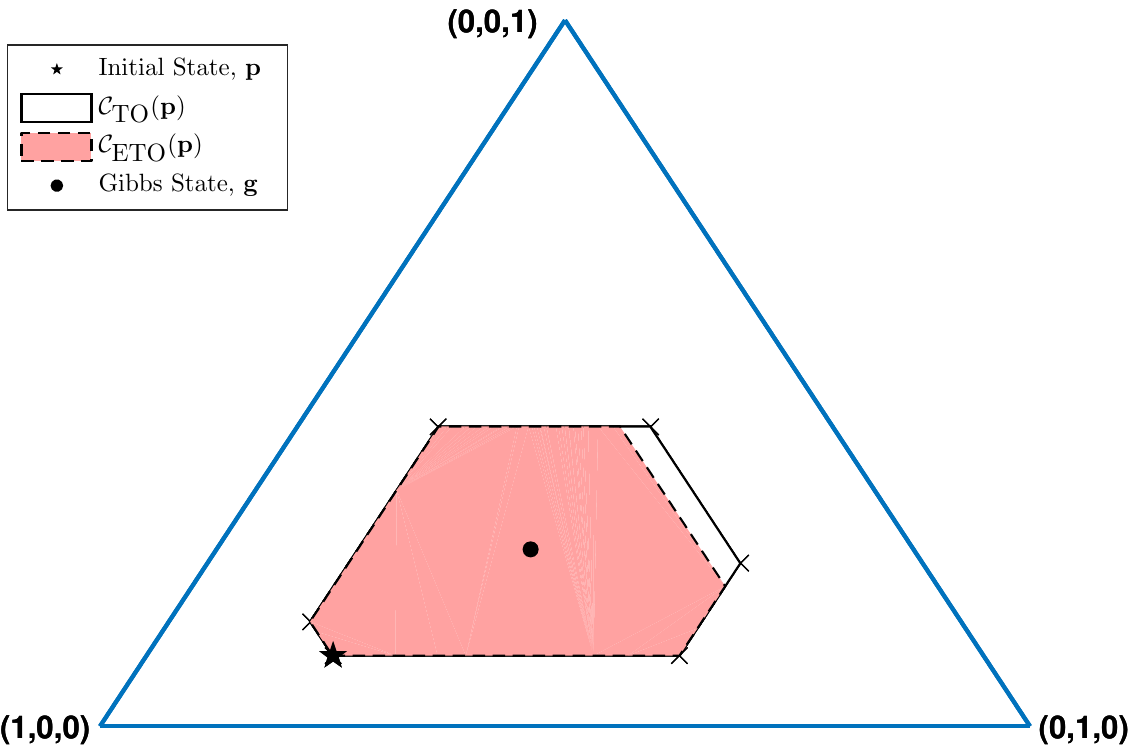}
\caption{\emph{Comparison of the ETO and TO cones:} Here we plot $\mathcal{C}_{\rm ETO}(\v{p})$ and $\mathcal{C}_{\rm TO}(\v{p})$ for a 3-level system in state $\v{p}$ and with associated energy levels $\v{E}$ on the 3 outcome probability simplex. Note the region of states accessible using TO but not by ETO. Here $\v{p}=\left(0.7,0.2,0.1\right)$, $\beta \v{E}=\left(0,0.2,0.5\right)$.}
\label{fig:ETO+TO_Cone}
\end{figure}

We complete the study of the main aspects of the resource theory of ETOs by addressing some thermodynamically relevant questions within the new framework: what is the maximum amount of work that can be extracted from a state? Conversely, how much work is necessary to prepare a state? We will show that there is a gap between what can be achieved with TO and ETOs. As a by-product, this shows that there will also be gaps between the predictions of the resource theory of thermal operations and physical models in which only ETOs are allowed.

\subsection{Work in the theory of ETOs} \label{ssec:work}

Under thermal operations, if it is not possible to convert $\v{p}$ into $\v{q}$ then an additional resource, work, may have to be supplied to achieve the conversion. Similarly, if $\v{p}$ can be converted into $\v{q}$, it may be possible to extract work. Within the resource theory approach, the \emph{deterministic} work can be regarded as the minimal amount of work that must be supplied to guarantee a transformation or the maximal amount of work that can be guaranteed to be extracted in converting one state into another. In this section, we investigate these concepts for ETOs and highlight the differences with the theory of thermal operations.

\begin{defn}[Deterministic Work under ETOs]
	\label{defn:deterministicstrictly}
	Given two probability distributions $\v{p}$ and $\v{q}$ with Hamiltonian $H_S$ and a 2-level system (called a \emph{wit} for short), with Hamiltonian $H_W$ which has energy levels $E^W_0=0$ and $E^W_1=w$, the \emph{work of transition} under ETO, $W^{\textrm{ETO}}_{\v{p}\rightarrow\v{q}}$, is the highest value of $w$ such that:
	\begin{equation} \label{eq:detwork}
		\v{p}\otimes \ketbra{0}{0}
		\etho
		\v{q}\otimes \ketbra{1}{1}.
	\end{equation}
	If $W^{\textrm{ETO}}_{\v{p}\rightarrow\v{q}}$ is positive, work is extracted in converting $\v{p}$ into  $\v{q}$, while if it is negative, work must be supplied to perform the transformation.
\end{defn}

This notion of deterministic work, which is stored in the extra 2-level wit system, is very common within the resource theory framework \cite{ horodecki2013fundamental,brandao2013second,aberg2013truly}. The main difference with other notions of work, such as those involving applying arbitrary unitaries to systems and the notion of passive states \cite{allahverdyan2004maximal} are: (a) only energy-preserving unitaries are allowed, (b) the unitaries can be applied to both system and heat bath rather than just system and (c) as the work is stored in a 2-level system rather than a (often implicit) weight-like device, it is a single-shot quantity, achieved in every run of an extraction protocol, rather than an average yield.

As a small caveat, note that the zero error statement of Eq.~\eqref{eq:detwork} is unobservable, and we can instead ask the more physically relevant question as to whether a given $w$ can be achieved up to an arbitrarily small, but potentially nonzero, error. I.e., is there a sequence of states \mbox{$(\v{q}\otimes \ketbra{1}{1})_\epsilon$} $\epsilon$-close to $\v{q}\otimes \ketbra{1}{1}$ and such that
\begin{equation} \label{eq:ETO smoothed}
	\v{p}\otimes\ketbra{0}{0}\etho (\v{q}\otimes \ketbra{1}{1})_\epsilon,
\end{equation} 
for every $\epsilon >0$? If this is the case, we will write \mbox{$W^{\textrm{ETO}}_{\v{p}\rightarrow\v{q}} \geq w$}. 

Let us denote the set of joint energy levels of system and wit by $\left\{\tilde{E}_i\right\}_{i=0}^{2d-1}$. These neatly divide into two sets. Firstly, let $A_0=\left\{\tilde{E}_i\right\}_{i=0}^{d-1}$ where $\tilde{E}_i=E_i$. These are associated with states for which the wit is in the state $\ket{0}$. Secondly, let $A_w=\left\{\tilde{E}_i\right\}_{i=d}^{2d-1}$ where $\tilde{E}_i=E_{i-d}+w$. These are associated with states for which the wit is in the state $\ket{1}$.

A crucial tool for analysing deterministic work under ETOs is given by the following Lemma:
\begin{lem} \label{lem:inj}
	Given distributions $\v{p}$ and $\v{q}$, if \mbox{$W^{\textrm{ETO}}_{\v{p}\rightarrow\v{q}} \geq w$}, 
	then there is a wit with energy gap $w$ and an injective map $\Phi$ between elements of $\tilde{E}_i\in A_0$ such that $p_i> 0$ and $A_w$ such that \mbox{$\Phi\left(\tilde{E}_i\right)\leq \tilde{E}_i$}.
\end{lem}
\begin{proof}
	Let us fix $w$. Suppose that with respect to this energy gap we have:
	\begin{equation*}
		\v{p}\otimes \ketbra{0}{0}
		\etho
		\v{q}\otimes \ketbra{1}{1},
	\end{equation*}
	and consider the convex combination of sequences of ETOs that implements this transformation. The initial state has support only on energy levels contained in $A_0$ while the final state has support only on energy levels contained in $A_w$. Hence, every sequence of ETOs within the convex combination must take $\v{p}\otimes \ketbra{0}{0}$ to a state of the form $\v{q}'\otimes \ketbra{1}{1}$. In what follows, we will consider one such sequence and use it to construct the required injection. We denote this sequence by $\pi$ and note that:
	\begin{align*}
		\v{q}'\otimes \ketbra{1}{1} & =   \pi\left(\v{p}\otimes\ketbra{0}{0}\right) \\ & = \left(G^{\left(i_n,j_n\right)}\dots G^{\left(i_1,j_1\right)}\right)\left(\v{p}\otimes\ketbra{0}{0}\right)
	\end{align*}
	where $G^{\left(i,j\right)}$ denotes an ETO acting non-trivially on energy levels $\tilde{E}_i$ and $\tilde{E}_j$.
	
	We now characterise the scenarios where an ETO acting on energy levels $\tilde{E}_i$ and $\tilde{E}_j$ maps the associated occupation probabilities $\left(r_i,r_j\right)$ such that $r_i>0$ to $\left(r'_i,r'_j\right)$ with $r'_i=0$. The occupation probabilities are related via: 
	\begin{equation}
		\begin{pmatrix}
			r'_i\\
			r'_j
		\end{pmatrix}
		=
		\begin{pmatrix}
			G_{i|i} & G_{i|j}\\
			G_{j|i} & G_{j|j}
		\end{pmatrix}
		\begin{pmatrix}
			r_i\\
			r_j
		\end{pmatrix},
	\end{equation}
	which implies that we need both $G_{i|i}=0$ and $r_j=0$. If $G_{i|i}=0$, then to achieve Gibbs-stochasticity we need that $G_{i|j} e^{-\beta \tilde{E}_j}=e^{-\beta \tilde{E}_i}$ which, together with $G_{i|j}\leq 1$, implies that it must hold that $\tilde{E}_i\geq\tilde{E}_j$.
	
	In light of this, we now modify our protocol $\pi$ as follows:
	\begin{enumerate}
		\item Without loss of generality assume that \mbox{$\tilde{E}_i\geq\tilde{E}_j$}. First, remove all of the $G^{\left(i,j\right)}$ such that \mbox{$G_{i|i}>0$}. Note that the resulting protocol still maps \mbox{$\v{p}\otimes\ketbra{0}{0}$} to a state of the form \mbox{$\hat{\v{q}}\otimes\ketbra{1}{1}$} as we have not removed any of the ETOs that empty population from a given energy level and in particular all population must eventually leave $A_0$.
		\item Next, remove all of the $G^{\left(i,j\right)}$ that are applied to intermediate states in the protocol such that $r_j>0$. Again, the resulting protocol still maps $\v{p}\otimes\ketbra{0}{0}$ to a state of the form $\hat{\v{q}}\otimes\ketbra{1}{1}$ as we have not removed any of the ETOs that empty population from a given energy level.
		\item Finally, remove all $G^{\left(i,j\right)}$ that are applied to intermediate states in the protocol such that \mbox{$r_i=r_j=0$} (as these operations now act trivially).
	\end{enumerate}

	We call the resulting sequence $\pi'$ and note that:

		\begin{equation*}
		\pi'\left(\v{p}\otimes\ketbra{0}{0}\right)=\hat{\v{q}}\otimes\ketbra{1}{1},
		\end{equation*}
	for some state $\hat{\v{q}}$ which potentially differs from both $\v{q}$ and $\v{q}'$. That the resulting state has this form, follows from the fact that the sequence of ETOs in $\pi'$ still contains the operations which empty the energy levels in $A_0$.
	
	The effect of each ETO in $\pi'$ is to swap the occupation probability of an occupied energy level $\tilde{E}_i$ with an unoccupied energy level $\tilde{E}_j$. In addition, each pair of energy levels is such that $\tilde{E}_i\geq \tilde{E}_j$. By combining each of these swaps to construct a permutation, we hence find an injection  $\Phi$ between elements of $A_0$ with $p_i> 0$ and $A_w$ such that \mbox{$\Phi\left(\tilde{E}_i\right)\leq \tilde{E}_i$}.
\end{proof}

As an example to illustrate the construction of such an injection, consider the two states $\v{p} = \left( 0, p_1, p_2\right)$ and $\v{q} = \left(p_1,p_2,0\right)$ with Hamiltonian \mbox{$H_S=\hbar \omega( \ketbra{1}{1} + 2 \ketbra{2}{2})$}. Suppose we wish to extract \mbox{$w=\hbar \omega$} in transforming $\v{p}$ into $\v{q}$. It is easy to see that this can be done by performing a sequence of two $\beta$-swaps. First, a $\beta$-swap is performed between the energy levels $\tilde{E}_1$ and $\tilde{E}_3$. Next, a $\beta$-swap is performed between the energy levels $\tilde{E}_2$ and $\tilde{E}_4$. The injection $\Phi$ is thus given by $\Phi\left(\tilde{E}_1\right) = \tilde{E}_3$, $\Phi\left(\tilde{E}_2\right) = \tilde{E}_4$.

\subsection{Unbounded work of formation}

Note that finding such an injection is a necessary, but not sufficient condition for being able to achieve $W^{\textrm{ETO}}_{\v{p}\rightarrow\v{q}} \geq w$. Indeed, while an injection satisfying the required properties will always exist for large negative values of $w$, there are state transformations for which supplying no finite amount of work enables the transformation, i.e. $W^{\textrm{ETO}}_{\v{p}\rightarrow\v{q}}=-\infty$. This is in stark contrast to the case of thermal operations where, provided the target state is block-diagonal in the energy basis, providing enough work will always make the transition possible, i.e. $W^{\textrm{ETO}}_{\v{p}\rightarrow\v{q}}>-\infty$.

Prime examples of this occur for a special case of deterministic work processes: the \emph{work of formation}. If $\v{g}$ denotes the probability distribution which is thermal with respect to $H_S$ at the fixed temperature $\beta$, then the work of formation of $\v{p}$ under ETO is the minimum amount of work required to deterministically create $\v{p}$ from the thermal state: $W^{\textrm{ETO}}_{\textrm{form}}\left(\v{p}\right)=W^{\textrm{ETO}}_{\v{g}\rightarrow \v{p}}$. If $\v{p}$ does not have full support, then Lemma \ref{lem:supp} implies the transformation is impossible regardless of the value of work chosen. Further examples that do not rely on arguments based on the size of the support of $\v{p}$ can also be constructed (see Appendix~\ref{appendix:extracounterexample}).

\subsection{Gap in work extraction}

The \emph{work of distillation} of $\v{p}$ under ETOs is the maximum amount of work that can be deterministically extracted from $\v{p}$ and is given by $W^{\textrm{ETO}}_{\textrm{distil}}\left(\v{p}\right)=W^{\textrm{ETO}}_{\v{p}\rightarrow \v{g}}$. A canonical toy example of work distillation is given by \emph{Szilard engines}. In its most basic form, this consists of a 2-level system with degenerate Hamiltonian $H_S\propto \iden$ and initially in state $\v{p}=\left(1,0\right)$. For such a setup, $W^{\textrm{TO}}_{\textrm{distil}}\left(\v{p}\right) = \beta^{-1} \ln 2$. However, under ETOs, for any value of $w>0$ it is not possible to find the injection required for work extraction as detailed in Lemma \ref{lem:inj} and hence $W^{\textrm{ETO}}_{\textrm{distil}}\left(\v{p}\right)=0$. This indicates that the resource theory of ETOs cannot exploit purity to extract work to the same extent as thermal operations.

This discrepancy between work distillation under TOs and ETOs is in fact completely general, as captured in the following theorem:
\begin{thm}
	Given a probability distribution $\v{p}$ and associated Hamiltonian $H_S$, if $W_{\textrm{distil}}^{\textrm{TO}}\left(\v{p}\right)>0$, then there exists a positive constant $C$ bounded away from zero such that:
	\begin{equation*}
	W_{\textrm{distil}}^{\textrm{ETO}}\left(\v{p}\right)+C\leq W_{\textrm{distil}}^{\textrm{TO}}\left(\v{p}\right).
	\end{equation*}
\end{thm}
\begin{proof}
Without loss of generality, let $\v{p}$ be ordered such that $p_i>0$ for \mbox{$i\in\left\{0,\dots,m-1\right\}$} and $p_i=0$ for \mbox{$i\in\left\{m,\dots,d-1\right\}$}. From the results of Ref.~\cite{horodecki2013fundamental} it follows that $W_{\textrm{distil}}^{\textrm{TO}}\left(\v{p}\right)$ satisfies:
\begin{equation} \label{eq:TO distil}
	e^{-\beta W_{\textrm{distil}}^{\textrm{TO}}\left(\v{p}\right)} Z_S=\sum_{i=0}^{m-1} e^{-\beta E_i}.
\end{equation}
For $W_{\textrm{distil}}^{\textrm{TO}}\left(\v{p}\right)>0$ to hold, we must have $m<d$.

	From Lemma~\ref{lem:inj}, to distil work $W_{\textrm{distil}}^{\textrm{ETO}}\left(\v{p}\right)$ we must be able to find an injection $\Phi$ such that \mbox{$\Phi(E_i):=E'_i+W_{\textrm{distil}}^{\textrm{ETO}}\left(\v{p}\right)\leq E_i$} for $i\in\left\{0,\dots,m-1\right\}$. 
	We thus have:
	\begin{equation*}
	\sum_{i=0}^{m-1} e^{-\beta E_i} \leq \sum_{i=0}^{m-1} e^{-\beta E'_i} e^{-\beta W_{\textrm{distil}}^{\textrm{ETO}}\left(\v{p}\right)}.
	\end{equation*}
	From Eq.~\eqref{eq:TO distil},
	\begin{equation*}
	W_{\textrm{distil}}^{\textrm{TO}}\left(\v{p}\right) \geq W_{\textrm{distil}}^{\textrm{ETO}}\left(\v{p}\right) -\frac{1}{\beta}\ln \sum_{i=0}^{m-1} \frac{e^{-\beta E'_i}}{Z_S}.
	\end{equation*}
	Set $C:=-\frac{1}{\beta}\ln \sum_{i=0}^{m-1} \frac{e^{-\beta E'_i}}{Z_S}$. The result follows from the fact that $C>0$, since $\{E'_i\}_{i=0}^{d-1}$ is a relabeling of the energies $\{E_i\}_{i=0}^{d-1}$ and $m<d$.
\end{proof}

\subsection{Quantum coherence and ETOs}
\label{sec:coherence}

So far we have focused on population dynamics. We could do this because under any ETO (in fact, any thermal operation) the evolution of population and coherence among energy eigenstates decouples (see Eq.~\eqref{eq:quantummapstostochasticmatrix}). However it is natural to wonder how the two theories compare in relation to their ability to process quantum coherence. 

In this subsection we consider the problem in its full generality: ETOs and thermal operations are two different sets of quantum channels (and not simply stochastic maps, as in the rest of this paper). How do they compare in terms of the coherent evolutions that they can induce?

First of all, it is important to clarify that ETOs do not coincide with the set of quantum operations that preserve the Gibbs state and act non-trivially only on two level subsystems (i.e., with the set of 2-level Gibbs-preserving operations). The simplest way to see this is that 2-level Gibbs-preserving channels can generate coherence from an initially incoherent state, while thermal operations, and hence ETOs, do not \cite{faist2015gibbs}. The two sets only coincide classically.

To tackle the issue of quantum coherence, we need to introduce the concept of \emph{modes of coherence}. Given a quantum state $\rho$, one can decompose it into parts that transform in the same way under time-translations \cite{marvian2014modes}. If $H_S$ has eigenvalues $\hbar \omega_i$, then
\begin{equation}
\rho = \sum_{\Omega} \rho^{(\Omega)}, \quad e^{-i H_S t} \rho^{(\Omega)} e^{i H_S t} = e^{-i \hbar \Omega t} \rho^{(\Omega)},
\end{equation}  
where the set of $\{\Omega\}$ is the set of all transition frequencies within $H_S$ (also called the Bohr spectrum of $H_S$, this is the set of all distinct differences $\omega_i - \omega_j$). Each $\Omega$ defines a so-called mode of coherence. If $\mathcal{E}$ is a thermal operation, coherence transformations can take place only within each mode, in the sense that 
\begin{equation}
\label{eq:covarianceproperty}
\mathcal{E}(\rho) = \sigma \Rightarrow \mathcal{E}(\rho^{(\Omega)}) = \sigma^{(\Omega)}, \quad \textrm{for each \;} \Omega,
\end{equation}
which generalises Eq.~\eqref{eq:quantummapstostochasticmatrix} from \mbox{$\Omega = 0$} to every $\Omega$ in the Bohr spectrum. Quantum maps satisfying Eq.~\eqref{eq:covarianceproperty} are called \emph{time-translation covariant}. 

Quantum coherence can be transported within each mode under thermal operations, sometimes perfectly \cite{lostaglio2015quantum}. On the other hand, we now show that coherence transport cannot take place under ETOs. Denoting $\rho_{xy} = \bra{x}\rho\ket{y}$,
\begin{lem}
	Let $\sigma = \mathcal{E}(\rho)$, with $\mathcal{E}$ an elementary thermal operation acting on energy levels $\omega_{x'}$ and $\omega_{y'}$, with $\omega_{x'} > \omega_{y'}$. Then
	\begin{equation}
	|\sigma_{x'y'}| \leq \sqrt{G_{x'|x'}G_{y'|y'}} |\rho_{x'y'}|.
	\end{equation}
\end{lem} 
\begin{proof}
	Thermal operations are time-translation covariant maps \cite{lostaglio2015description}. ETOs are a subset of them, so they are as well. Theorem~2 of Ref.~\cite{lostaglio2017markovian} then immediately implies that,
	\begin{equation}
	\label{eq:boundcoherence}
	|\sigma_{x'y'}| \leq \sum_{\substack{x,y: \\ \omega_x-\omega_y=\omega_{x'}-\omega_{y'}}} \sqrt{G_{x'|x} G_{y'|y}} |\rho_{xy}|,
	\end{equation} 
	where the sum is restricted to all indexes such that $\omega_x - \omega_y = \omega_{x'} - \omega_{y'}$ and, as before, \mbox{$G_{k'|k} = \bra{k'}\mathcal{E}(\ketbra{k}{k})\ket{k'}$}. 
	
 Consider now any term in the above sum in which $\omega_{x'} \neq \omega_{x}$. Since \mbox{$\omega_{y'} - \omega_y = \omega_{x'} - \omega_x$}, it follows that \mbox{$\omega_y \neq \omega_{y'}$} (so there are no terms in which $\omega_{x'} \neq \omega_{x}$ but $\omega_{y'} = \omega_{y}$). From \mbox{$\omega_{x'} \neq \omega_{y'}$} one gets \mbox{$\omega_x \neq \omega_y$}. Furthermore, \begin{align}
\omega_{x'} - \omega_y &=& \omega_x - \omega_y + \omega_{y'} - \omega_y \\
\omega_{y'} - \omega_x &=& - (\omega_x - \omega_y) + \omega_{x'} - \omega_x.
	\end{align}
	Since $\omega_{y'} - \omega_y = \omega_{x'}- \omega_x$, it follows that one could have $\omega_{x'} = \omega_y$ or $\omega_{y'} = \omega_x$, but not both. This shows, for every term with $\omega_{x'} \neq \omega_x$ or $\omega_{y'} \neq \omega_y$, that at most two of the four levels $\omega_{x'}$, $\omega_{x}$, $\omega_{y'}$, $\omega_y$ have the same energy. Hence, three distinct energy levels are involved in the corresponding pre-factor $\sqrt{G_{x'|x} G_{y'|y}}$. ETOs, on the other hand, can only act non-trivially on two energy levels at any single time. Hence, either $G_{x'|x} = 0$ or $G_{y'|y} = 0$, so every such term is zero. Eq.~\eqref{eq:boundcoherence} follows.
\end{proof}
The previous lemma implies that there cannot be any transport of coherence within a mode. For example, given a qutrit system with \mbox{$H_S = \hbar \omega (\ketbra{1}{1} + 2\ketbra{2}{2})$} (the ground state having zero energy), one cannot move coherence from $\ketbra{0}{1}$ into $\ketbra{1}{2}$ using ETOs, even though these are part of the same mode $\Omega = \omega$. This shows that ETOs are more limited than thermal operations, not only from the point of view of population dynamics, but also in terms of coherence processing. Another interesting consequence of the previous lemma is that $\beta$-swaps destroy all quantum coherence in the levels on which they act, since they have either \mbox{$G_{x'|x'} = 0$} or \mbox{$G_{y'|y'} = 0$}. 

A superset of ETOs that can be easily characterised is that of \emph{enhanced ETOs}, which are defined as enhanced thermal operations acting on 2 energy levels. Recall that an enhanced thermal operation is a quantum channel $\mathcal{E}$ such that \cite{cwiklinski2015limitations}
\begin{enumerate}
	\item $\mathcal{E}(e^{-\beta H_S}) = e^{-\beta H_S}$,  \quad (Gibbs-preservation)
\item Eq.~\eqref{eq:covarianceproperty} is satisfied,  \quad (Time-translation covariance)
\end{enumerate}
There are enhanced ETOs achieving the bound of the previous lemma. Their Kraus decomposition is given as (taking $\omega_{x'}>\omega_{y'}$)
\begin{small}
\begin{align*}
K_1 = & \sqrt{e^{-\beta \hbar \omega_{x'y'}}G_{y'|x'}} \ketbra{x'}{y'}, \quad K_{-1} = \sqrt{G_{y'|x'}} \ketbra{y'}{x'}, \\
K_0 = & \sqrt{1-G_{y'|x'}e^{-\beta \hbar \omega_{x'y'}}} \ketbra{y'}{y'} + \sqrt{1-G_{y'|x'}} \ketbra{x'}{x'}.
\end{align*}
\end{small}
In fact, one can deduce from Ref.~\cite{cwiklinski2015limitations} that ETOs achieving the bound exist as well.

It is worth noticing that the parallel we drew in Section~\ref{sec:collision} between ETOs, collision models and Davies maps extends to the quantum case. Taking $G_{k'|k} = T/n$ (where $T>0$ denotes the total time and $n$ the number of collisions) one obtains in the \mbox{$n \rightarrow \infty$} limit of infinitely many weak collisions the standard \cite{roga2010davies} exponential damping of each coherence term, \mbox{$|\sigma_{x'y'}|:=\rho_{x'y'}(T) \leq e^{-\xi T} |\rho_{x'y'}|$}, $\xi > 0$.  

\section{Thermodynamic resource theories at finite temperature}

So far, we have introduced and analysed ETOs as a suitable set of allowed transformations, distinct from TOs and related to physically relevant models. We now broaden the discussion to more general thermodynamic resource theories. We discuss how, in the infinite temperature limit, many different thermodynamic theories ``flow'' toward a universal model, characterised by majorisation theory. On the one hand, we can reinterpret the results of the previous sections as showing that this universality is lost at finite temperature. On the other hand, we show that some of the techniques developed in this manuscript can be applied more generally and tailored very specifically to the experimental constraints at hand. 

\subsection{Lack of universality at finite temperature}
\label{sec:lack}

What set of operations can be performed at no work cost, if we require compatibility with the second law of thermodynamics? Each (state-independent) choice defines a \emph{thermodynamic resource theory}.\footnote{A framework not included in this discussion is, for example, catalysis~\cite{brandao2013second}. Catalysts allow one to perform some non Gibbs-stochastic maps on some non thermal states, as can be deduced from the fact that they present weaker constraints than thermo-majorisation \cite{brandao2013second}.}
	
As mentioned before, thermal operations realise the full set of Gibbs-preserving maps on the energy levels' populations. In terms of allowed population dynamics, this is arguably the largest thermodynamic resource theory, since any other stochastic process would allow the extraction of work from a single heat bath. This follows from the fact that if a map $M$ with $M\v{g} = \v{x}$ and $\v{x} \neq \v{g}$ is allowed ($\v{g}$ being the thermal Gibbs state of Eq.~\eqref{eq:gibbstate}), then we can create many copies of $\v{x}$ from many copies of $\v{g}$; Theorem~1 of Ref.~\cite{brandao2011resource} then implies that we could raise a weight using a single bath, in disagreement with the second law of thermodynamics. A much more restricted theory is one in which only partial level thermalisations (PLTs) on pairs of levels are allowed (see Eq.~\eqref{eq:partiallevelthermalisation}). At infinite temperature, the classical result of Theorem~\ref{thm:muirhead} immediately implies the following universality result:
\begin{lem}
	In the limit $T \rightarrow \infty$, every thermodynamic theory that allows for PLTs on pairs of levels and does not extract work from a single heat bath is governed by the same laws at the classical level. In other words, if $A$ and $B$ are two such theories and $\v{p}$, $\v{q}$ are any two states
	\begin{equation*}
\v{p} \stackrel{\textsc{\tiny A}}{\rightarrow} \v{q} \quad \Leftrightarrow  \quad \v{p}   \stackrel{\textsc{\tiny B}}{\rightarrow} \v{q} 
	\end{equation*}
\end{lem}

Furthermore, well-known results on doubly-stochastic maps (\cite{marshall2010inequalities}, Chapter 2) imply that at infinite temperature a transformation is possible if and only if $\v{p}$ majorises $\v{q}$. This remarkable degree of independence of the thermodynamic constraints from the specific choice of the set of allowed free operations is lost at finite temperature, as shown by the counterexample of Sec.~\ref{sec:ETO theory}. Also, universality does not hold for transformations involving quantum coherence since, as mentioned in Sec.~\ref{sec:coherence}, Gibbs-preserving quantum operations and thermal operations allow different sets of transformations. This lack of universality implies that to obtain tighter constraints on the allowed transformations we need to develop models that are more closely connected to the relevant experimental conditions.

For example, we can observe that the construction of Theorem~\ref{thm:extremal} holds for any restriction of the set of ETOs, provided that PLTs are allowed. In the next section we explicitly consider the example of the Jaynes-Cummings model in RWA, and then consider a generalisation that characterises a large class of thermodynamic theories given the knowledge of the set of extremal maps.

\subsection{The resource theory of Jaynes-Cummings operations}

We have demonstrated that the theory of ETOs combines appealing theoretical properties with the ability to reproduce, within a large set of parameters, 2-level Jaynes-Cummings interactions. However, for high dimensional systems, due to the need of applying long sequences of operations, the differences between the ETOs and the Jaynes-Cummings model will become important. Hence, we now go a step further and define a restricted resource theory, contained within ETOs, that coincides \emph{exactly} with the Jaynes-Cummings model.

More precisely, we are assuming that one can perform the operations described in Section~\ref{sec:JC}, i.e. all stochastic matrices
\begin{equation}
	J^{(i,j)}(s)=\begin{pmatrix}
		1- G_{j|i}(s) e^{-\beta \hbar \omega_{ji}}& G_{j|i}(s)  \\
		G_{j|i}(s) e^{-\beta \hbar \omega_{ij}}  & 1-G_{j|i}(s) 
	\end{pmatrix},
\end{equation} 
acting on any pair of levels $i,j$, $E_i \geq E_j$, and satisfying Eq.~\eqref{eq:JCtransitions} (recall that $s$ is the normalised interaction time, $s = gt/\hbar$).  Sequential composition and convex combinations are allowed. With obvious notation,
	\begin{defn}[JC cone]
		Given an initial state $\v{p}$, let $\mathcal{C}_{\rm JC}(\v{p})$ be given by all $\v{q}$ such that $\v{p} \jc \v{q}$.
	\end{defn}

As in ETOs, any operation among two given levels is described by a single parameter, $G_{j|i}(s)$. If $s$ is limited by some $s^{\left(i,j\right)}_{\rm max}$ (possibly dependent on the levels $i$, $j$ involved), then $G_{j|i}(s)$ can be numerically optimised to find its maximum within $\left[0, s^{\left(i,j\right)}_{\rm max}\right]$. If the interaction time $s$ is not limited, one needs to find bounds independent of $s$, for example the upper and lower bounds given at the end of Sec.~\ref{sec:JC}. In every case, we are left with a value (or interval) for the greatest relaxation probability $G_{\rm max}(\beta \hbar \omega_{ij})$ allowed in the model for any given transition. The only assumption we will make is that the theory allows to perform PLTs between any pair of energy levels, i.e. $G_{\rm max}(\beta \hbar \omega_{ij}) > 1/(1+e^{-\beta \hbar \omega_{ij}})$ for all $i$,$j$.\footnote{The reason for the strict inequality is a technical one. The second part of the proof of Theorem \ref{thm:extremal} required that we could decompose PLTs into non-trivial convex combinations of extremal maps and identity transformations. If the extremal map were a PLT (the inequality was not strict), this would not be possible.}

We now proceed in constructing $\mathcal{C}_{\rm JC}(\v{p})$. This can be done immediately from Theorem~\ref{thm:extremal}, since it holds unchanged within this restricted theory, with the caveat of substituting $\beta$-swaps with the relevant extremal maps:
\begin{equation*}
J^{(i,j)}=\begin{pmatrix}
1- G_{\rm max}(\beta \hbar \omega_{ij}) e^{-\beta \hbar \omega_{ij}}& G_{\rm max}(\beta \hbar \omega_{ij})  \\
G_{\rm max}(\beta \hbar \omega_{ij}) e^{-\beta \hbar \omega_{ij}}  & 1-G_{\rm max}(\beta \hbar \omega_{ij}) 
\end{pmatrix}.
\end{equation*}
If we only know that  $G_{\rm max}(\beta \hbar \omega_{ij}) \in [G^{<}_{ij}, G^{>}_{ij}]$, as in our Jaynes-Cummings example if we do not enforce any limitations on $s$, we can still build an outer cone $\mathcal{C}^{>}_{\rm JC}(\v{p})$ and an inner cone $\mathcal{C}^{<}_{\rm JC}(\v{p})$ satisfying \mbox{$
\mathcal{C}^{>}_{\rm JC}(\v{p}) \supset \mathcal{C}_{\rm JC}(\v{p}) \supset \mathcal{C}^{<}_{\rm JC}(\v{p})$}, simply using the lower or upper bounds of Fig.~\ref{fig:achievable_region}.  

\begin{figure}[t!] 
	\includegraphics[width=1\columnwidth]{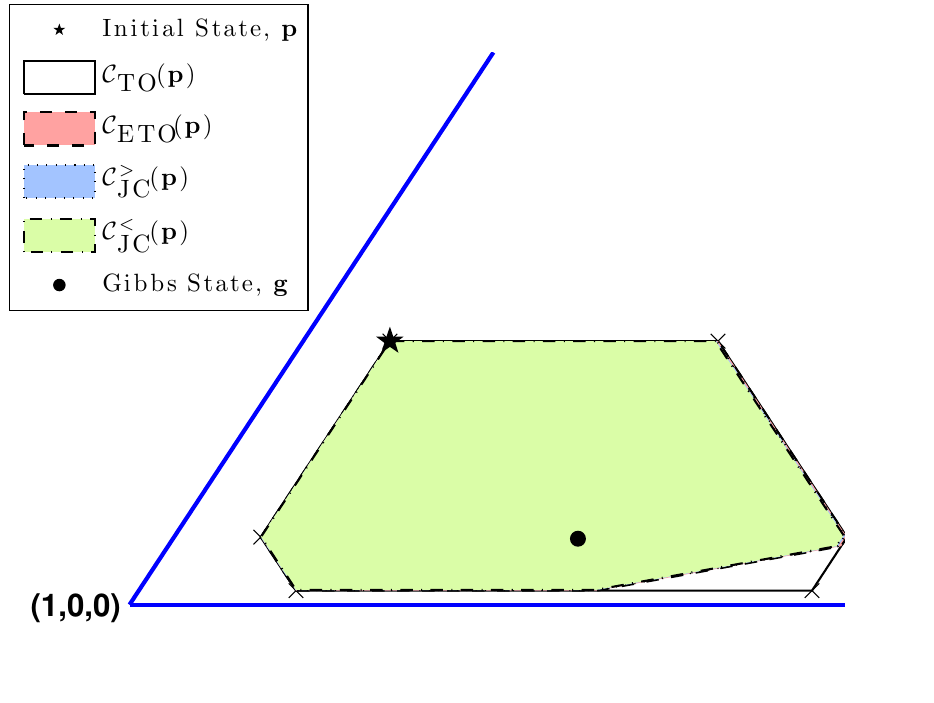}
	\caption{\emph{Comparison of the ETO and JC cones for $T=300K$:} Here we plot $\mathcal{C}_{\rm ETO}(\v{p})$, $\mathcal{C}_{\rm TO}(\v{p})$ and upper and lower bounds on $\mathcal{C}_{\rm JC}(\v{p})$ for a 3-level system in state $\v{p}$ and with associated energy levels $\v{E}$ on the 3 outcome probability simplex. Here 
		$\v{p}=\left(0.85,0.03,0.12\right)$, 
		$\v{E}/\hbar=\left(0,69 \, {\rm THz},130 \, {\rm THz}\right)$.}
	\label{fig:JC300K}
	\vspace{0.5cm}
	\includegraphics[width=1\columnwidth]{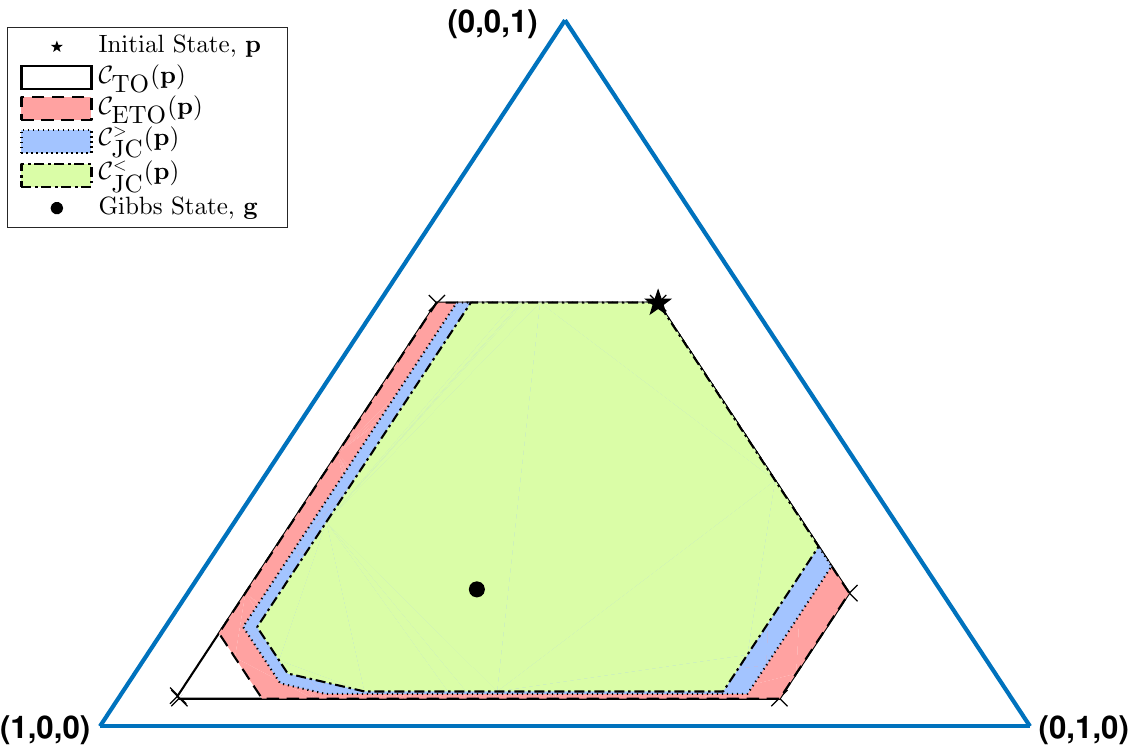}
	\caption{\emph{Comparison of the ETO and JC cones for $T=0.1K$:} Here we plot $\mathcal{C}_{\rm ETO}(\v{p})$, $\mathcal{C}_{\rm TO}(\v{p})$ and upper and lower bounds on $\mathcal{C}_{\rm JC}(\v{p})$ for a 3-level system in state $\v{p}$ and with associated energy levels $\v{E}$ on the 3 outcome probability simplex. Here $\v{p}=\left(0.1,0.3,0.6\right)$, $\v{E}/\hbar=\left(0,6.2 \, {\rm GHz},11.7 \, {\rm GHz} \right)$ . } 
	\label{fig:JCmiliK}
\end{figure}

This allow us to construct cones that faithfully reproduce the specific experimental constraints at hand in the Jaynes-Cummings model described above. Some results are presented in Figs.~\ref{fig:JC300K} and \ref{fig:JCmiliK}. Fig.~\ref{fig:JC300K} considers optical frequencies and room temperature; this is a regime in which, from the considerations of Sec.~\ref{sec:etophysicalmodels}, we expect ETOs to provide good constraints for the JC model. In fact, we can see that the allowed transitions are almost indistinguishable, i.e. $\mathcal{C}_{\rm JC}(\v{p}) \approx \mathcal{C}_{\rm ETO}(\v{p})$, while there is a clear distinction between $\mathcal{C}_{\rm ETO}(\v{p})$ and $\mathcal{C}_{\rm TO}(\v{p})$.
Fig.~\ref{fig:JCmiliK}, in contrast, considers microwave frequencies and a temperature of $0.1K$, a parameter regime in which we do \emph{not} expect ETOs to be a provide a faithful model. In fact, the correspondent cones are rather different, highlighting the lack of thermodynamic universality and the usefulness of the techniques developed here to build resource theories that fit the specific experimental conditions.  

\subsection{Restricted thermodynamic theories}

The previous considerations can be extended to more general restricted thermodynamic theories. As an example, Ref.~\cite{mazurek2017preparation} has identified the set of extremal Gibbs-preserving maps involving $3$-levels at any time. Then, one would like to construct the set of achievable states for the $3$-dimensional generalisation of ETOs, in which we allow all extremal 2-level and 3-level maps in the set of Gibbs-stochastic maps. More generally, one could consider theories in which every extremal map on $k$ energy levels is allowed, with $k=2$ corresponding to ETOs and $k=d$ thermal operations (see Ref.~\cite{mazurek2017preparation}). Another appealing restriction is given by ETOs augmented by a global PLT involving all energy levels at once or, in a more refined version, PLTs on any set of $k>2$ levels. Ultimately, the relevant set of restrictions will depend on the experimental setup. However, all of the above theories can be solved from the knowledge of the set of extremal maps using the following generalisation of Theorem~\ref{thm:extremal}:

\begin{thm}
	Let $A$ denote the allowed operations in a convex thermodynamic resource theory. Suppose $A$ contains, for every $i, j$, $E_i > E_j$ a map
	\begin{equation}
	A^{(i,j)}=\begin{pmatrix}
	1- G^{\rm max}_{j|i} e^{-\beta \hbar \omega_{ij}}& G^{\rm max}_{j|i}  \\
	G^{\rm max}_{j|i} e^{-\beta \hbar \omega_{ij}}  & 1-G^{\rm max}_{j|i} 
	\end{pmatrix},
	\end{equation}
	for some $G^{\rm max}_{j|i} > 1/(1+e^{-\beta \hbar \omega_{ij}})$ (in other words, 2-level PLTs are strictly contained).  
	
	If $A_i$ are the extremal maps in $A$, then $\v{p} \stackrel{\textsc{\tiny A}}{\rightarrow} \v{q}$ if and only if $\v{q}$ is in the convex hull of the following points:
		\begin{equation*}
		\label{eq:sequenceextremal}
		 \{A_{i_n} \cdots A_{i_1} \v{p}\},
		\end{equation*}
		where $n \leq d!$. 
\end{thm}
\begin{proof}
	The proof is a straightforward generalisation of that of Theorem~\ref{thm:extremal}. Firstly, we can prove that any extremal points of the set $\{\v{r}| \v{p} \stackrel{\textsc{\tiny A}}{\rightarrow} \v{r} \}$ must be achievable through a sequence of extremal maps in $A$, as in Eq.~\eqref{eq:sequenceextremal}. Secondly, if by contradiction we need $n>d!$ extremal maps, then some $\beta$-ordering will be repeated at least twice in the sequence; but then we can substitute part of the sequence with 2-level PLTs. Since by assumption these 2-level PLTs can be written as a non-trivial convex combination of extremal points in $A$, the rest of the proof of Theorem~\ref{thm:extremal} holds. 
\end{proof}
  
\section{Applications in cooling algorithms} \label{ap:HBAC}

We now explore one potential application of the framework presented in the previous sections, regarding the cooling of quantum states. The generation of highly pure, ``cold'', quantum states is one of the necessary resources in the implementation of quantum technologies \cite{divincenzo2000physical}. These are needed, for instance, as ancilla qubits in quantum error correction, or at the initial stage of many quantum algorithms. 

A potential method for generating these pure states is via \emph{algorithmic cooling}, which is a generic name for a set of schemes in which unitary operations are applied to a set of qubits in order to reduce the entropy of individual ones. It was found that such entropy extraction processes can be enhanced with the introduction of interactions with a heat bath \cite{boykin2002algorithmic}. These so called \emph{Heat Bath Algorithmic Cooling} (HBAC) protocols, can thus be thought of as sets of thermodynamical operations. For a review of HBAC in the context of NMR experiments, see Ref.~\cite{park2016heat}.

In such a scenario, we have a set of $n$ qubits with identical individual Hamiltonians $H_{S_i}=\frac{\Delta}{2} \sigma_Z \otimes \iden_{\backslash S_i}$, where \mbox{$\sigma_Z=\ketbra{1}{1}-\ketbra{0}{0}$}
is the Pauli-$Z$ operator, $\iden_{\backslash S_i}$ is the identity on the complementary subspace to qubit $i$ and \mbox{$i=0,..,n-1$}. Here $\Delta$ denotes the energy gap. The aim is to perform operations that increase the \emph{polarisation} of a target qubit, say $i=0$, as much as possible, taking advantage of the auxiliary qubits. Given a density matrix $\rho$, the polarisation of the target qubit is defined as
\begin{equation*}
\epsilon=\tr{}{\rho (\sigma_Z \otimes \iden_{\backslash S_0})}.
\end{equation*}
Note that in terms of $\epsilon$, the population of qubit $i=0$ can be written as
\begin{equation*}
\v{p}=\frac{1}{2}\left(1+\epsilon,1-\epsilon\right).
\end{equation*}

Given that these protocols involve thermodynamical operations on qubits, one can ask whether having access to the set of ETOs provides an enhancement to the performance of existing algorithms. We initiate the study of this question by analysing the recent cooling scheme introduced in Ref.~\cite{rodriguez2017heat}.

\begin{figure}[t!]
	\includegraphics[width=0.9\columnwidth]{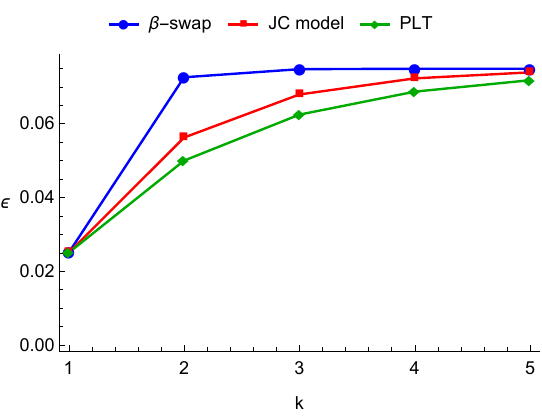}
	\caption{\emph{Comparison of 2 qubit HBAC protocols.} Here we show how the polarisation $\epsilon = \langle \sigma_Z \rangle$ of a qubit increases with the number of iterations $k$ in the HBAC protocol described in Section~\ref{ssec:HBAC 2 qubit}. We compare our $\beta$-swap protocol to the original PLT protocol presented in Ref.~\cite{rodriguez2017heat}. One sees that the new protocol reaches the asymptotic polarisation value more rapidly than the original. Also the JC model performs better, even though less remarkably so.}
	\label{fig:HBAC-JC}
\end{figure}

In this work, the authors use protocols containing PLTs to boost the performance of HBAC. Building on that idea, we now present a modified HBAC based on $\beta$-swaps, and explore how these improve the performance of the protocols. In the low polarisation limit we show that the asymptotic cooling of the HBAC protocol proposed in Ref.~\cite{rodriguez2017heat} can be achieved in a single step using $\beta$-swaps. Furthermore, we give a preliminary discussion of the cooling achievable within a Jaynes-Cummings model, that also improves on the HBAC based on PLTs although less significantly (see Fig.~\ref{fig:HBAC-JC}). While this provides another illustration of the applications of our framework, we leave a more in-depth study to further work.

\begin{figure}[t]
	\includegraphics[width=1\linewidth]{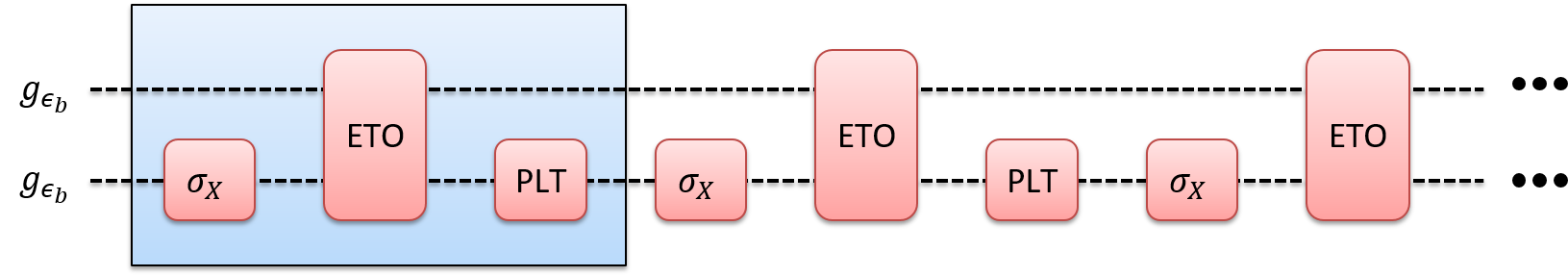}
	\caption{\emph{HBAC protocol for 2 qubits.} In this figure we present a generalisation of the protocol given in \cite[Fig.~4]{rodriguez2017heat}. While they consider a thermalisation step for the ETO, we will instead utilise a $\beta$-swap. The box contains one iteration of the protocol.}
	\label{fig:Alg2} 
\end{figure}

\subsection{Enhanced HBAC with two-qubit interactions} \label{ssec:HBAC 2 qubit}

The scheme presented in Ref.~\cite{rodriguez2017heat} involves a set of $n$ qubits and an iterative procedure in which thermalisation of joint energy eigenstates is performed. We begin by explaining the protocol for $n=2$, which is illustrated in Fig.~\ref{fig:Alg2}.

The scheme begins with two qubits: $S_0$ (target) and $S_1$ (auxiliary), in their thermal states \mbox{$\v{g}_{\epsilon_b}=\frac{1}{2}\left(1+\epsilon_b,1-\epsilon_b\right)$}, where $\epsilon_b$ denotes the polarisation of the Gibbs state and is related to the energy gap of the Hamiltonian via:
\begin{equation*}
\epsilon_b=\tanh{\left(\beta \frac{\Delta}{2}\right)}.
\end{equation*}
The protocol in Ref.~\cite{rodriguez2017heat} then runs as follows:
\begin{enumerate}
	\item Realise a population inversion on qubit $S_1$ via a Pauli-$X$ gate. The joint state of the two qubits is now athermal. 
	\item \label{step2} Apply a fully thermalising PLT between the joint energy levels $\ket{0}_{S_0}\ket{0}_{S_1}$ and $\ket{1}_{S_0}\ket{1}_{S_1}$. This is an ETO for which matrix acting on those two levels can be written as 
	\begin{equation*}
	T^{\left(00,11\right)}_{\textrm{therm}}	= \frac{1}{1+e^{-\beta 2 \Delta}} \begin{pmatrix}
	1 & 1 \\
	e^{-\beta 2 \Delta} & e^{-\beta 2 \Delta}
	\end{pmatrix},
	\end{equation*} 
	This can be seen from Eq.~\eqref{eq:pltmatrix}, since \mbox{$T^{\left(00,11\right)}_{\textrm{therm}}	:= T_{\lambda = 1/(1+e^{-\beta 2\Delta})}$} with $\hbar \omega = 2\Delta$ and $(i,j) = (00,11)$.
	\item Reset qubit $S_1$ to the initial thermal state via another fully thermalising PLT. In terms of the energy levels, this corresponds to applying two full thermalisations, $T^{\left(00,01\right)}_{\textrm{therm}}$ and $T^{\left(10,11\right)}_{\textrm{therm}}$.
	\item Repeat until the polarisation of qubit $S_0$ reaches a stationary value.
\end{enumerate}

For small initial polarisation $\epsilon_b$, it was shown that after $k$ iterations of the above protocol the polarisation of the target is given by \cite{rodriguez2017heat}
\begin{equation}\label{eq:lowpo2}
\epsilon_k^{n=2}=\left( 3-\frac{1}{2^{k-1}}\right) \epsilon_b.
\end{equation}
Hence, the polarisation triples in the limit of large $k$.

More generally, for any $\epsilon_b$, the polarisation after each run obeys the recursion relation
\begin{equation*}
\epsilon_{k+1}^{n=2}=\frac{\text{sech}(\beta \Delta)}{2}\left[\sinh{\left(\frac{3}{2} \beta \Delta\right)}\text{sech}\left(\beta \frac{\Delta}{2}\right)+\epsilon_{k}^{n=2}\right],
\end{equation*}
from this the asymptotic value $\epsilon_{\infty}^{n=2}=\tanh{\left(\frac{3}{2} \beta \Delta\right)}$ can be derived.

Here, we shall modify the above algorithm, substituting the fully thermalising PLT of step~\ref{step2} for a $\beta$-swap between the two energy levels:
\begin{equation*}
\beta^{\left(00,11\right)}=	\begin{pmatrix}
1-e^{-2\beta\Delta} & 1 \\
e^{-2\beta\Delta} & 0
\end{pmatrix}.
\end{equation*}
The first iteration of this algorithm creates
\begin{align*}
\v{g}_{\epsilon_b}^{\otimes 2}=&\frac{1}{4 \cosh^2{\left(\beta \frac{\Delta}{2}\right)}}\left(e^{\beta \Delta}, 1, 1, e^{-\beta \Delta}\right) \\
\stackrel{\text{\tiny$\sigma_X$}}{\longrightarrow}&\frac{1}{4 \cosh^2{\left(\beta \frac{\Delta}{2}\right)}}\left(1, e^{\beta \Delta}, e^{-\beta \Delta},1\right) \nonumber \\
\stackrel{\text{\tiny$\beta$-swap}}{\longrightarrow}& \frac{1}{4 \cosh^2{\left(\beta \frac{\Delta}{2}\right)}}\left(2-e^{-2 \beta \Delta}, e^{\beta \Delta}, e^{-\beta \Delta},e^{-2 \beta \Delta}\right) \nonumber\\
\stackrel{\text{\tiny PLT}}{\longrightarrow}& \frac{1}{4 \cosh^2{\left(\beta \frac{\Delta}{2}\right)}}\left(2-e^{-2 \beta \Delta}+ e^{\beta \Delta}, e^{-\beta \Delta}+ e^{-2 \beta \Delta}\right) \nonumber \\\nonumber &\qquad \qquad \qquad \qquad \qquad \otimes \frac{1}{1+e^{-\beta \Delta}}\left(1,e^{-\beta \Delta}\right)
\end{align*}

More generally, one obtains the following recursion relation for the polarisation after each run
\begin{equation*}
\tilde{\epsilon}_{k+1}^{n=2}=\frac{\tilde{\epsilon}_{k}^{n=2} \left(e^{2 \beta \Delta }-1\right)+e^{3\beta \Delta}-1}{e^{2\beta \Delta }+e^{3\beta \Delta }}.
\end{equation*}
This also converges to $\tanh{\left(\frac{3}{2} \beta \Delta\right)}$ in the limit of large $k$. A comparison of the rate at which the two protocols converge to this asymptotic value is given in Fig.~\ref{fig:HBAC-JC}.

If one focuses on the low polarisation regime, it can be shown from the above derivation that for the $\beta$-swap protocol
\begin{equation*}
\tilde{\epsilon}^{n=2}_{k=1}=3 \epsilon_b.
\end{equation*}
That is, the asymptotic value in the low polarisation regime is achieved after a \emph{single run} of the algorithm. This is in stark contrast to the algorithm using a thermalisation step where this polarisation is only achieved in the limit of large $k$. From the discussion in Sec.~\ref{sec:etophysicalmodels} one can infer that this enhancement is due to the fact that $\beta$-swaps are non-markovian processes.

\subsection{Implementation via Jaynes-Cumming model}

The previous results indicate that considering more general thermal operations than thermalisations in cooling protocols can improve performance dramatically. Of course, a more detailed investigation is necessary in order to assess the feasibility of the proposal within specific platforms and to fairly compare with other HBAC proposals. While this goes beyond the scope of the present investigation, a preliminary discussion of this topic constitutes a useful illustration of how the theoretical framework presented here can be used in practice to explore new interesting thermodynamic protocols.

For low initial polarisation (i.e. high temperatures, the regime in which our protocol at face value improves considerably over the proposal of Ref.~\cite{rodriguez2017heat}), the Jaynes-Cummings model can only realise poor approximations of $\beta$-swaps. Nevertheless, improvements can still be seen. As an example, assume one implements step~\ref{step2} of the cooling protocol through a Jaynes-Cummings model with $\beta \hbar \Delta = 0.05$ and a suitable coupling $g$ that may depend on the relevant experimental parameters. One then has $G_{1|0}(t)$ following Eq.~\eqref{eq:JCtransitions}. $G_{1|0}$ reaches a first local maximum quite quickly (at around $\bar{t}=1.5t_{th}$ with $t_{th}$ being the time necessary to implement a full thermalisation). At that point, one has $G_{1|0}(\bar{t}) \approx 0.656$. Despite this being very far from the probability required by a $\beta$-swap (i.e., $G_{1|0} \approx 1$) one can see the improvements over the PLT protocol, see Fig.~\ref{fig:HBAC-JC}. In fact, a more detailed analysis shows that after a single step the improvement on the HBAC protocol is linear in $G_{1|0}(\bar{t}) - G_{1|0}(t_{th})$, i.e. in the excess de-excitation probability as compared to the full thermalisation process. We only mention in passing that a variety of Jaynes-Cummings protocols can be obtained by imposing different time constraints, and this will depend on the details of the experimental proposal.

\subsection{Many qubit case}

\begin{figure}[t]
	\includegraphics[width=1\linewidth]{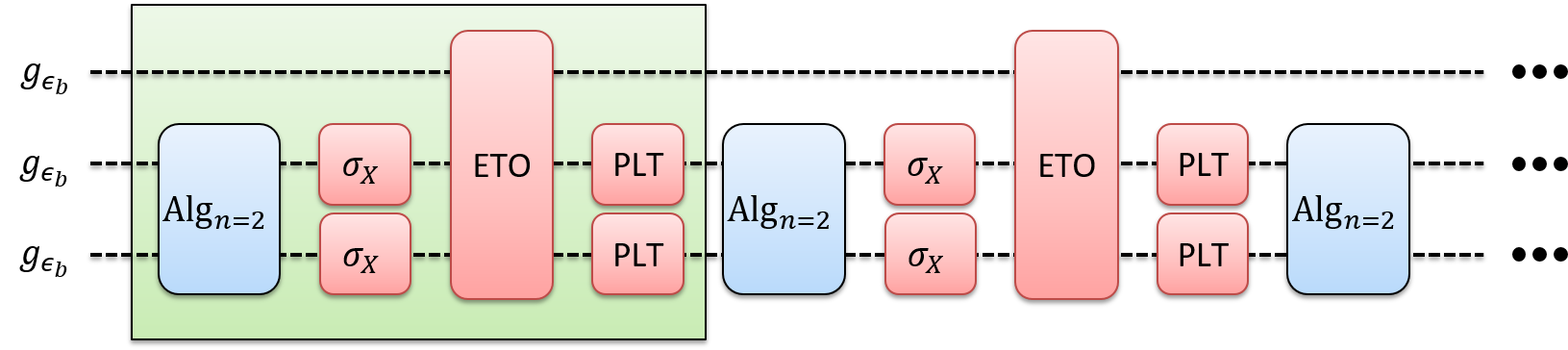}
	\caption{\emph{HBAC protocol for $3$ qubits.} In this figure we present a generalisation of the protocol given in \cite[Fig.~6]{rodriguez2017heat}. While they consider a thermalisation step for the ETO, we will instead utilise a $\beta$-swap. $\textrm{Alg}_{n=2}$ implements the algorithm from Fig.~\ref{fig:Alg2}. The box contains one iteration of the protocol.}
	\label{fig:Alg3} 
\end{figure}

The above algorithms can be extended to act on an arbitrary number of auxiliary qubits $S_1,..,S_{n-1}$, enhancing the final asymptotic polarisation for the target qubit $S_0$. It becomes an iterative algorithm, in the sense that each step on $n$ qubits requires an implementation of the algorithm for $n-1$ qubits as detailed in Fig.~\ref{fig:Alg3} for $n=3$.

\begin{figure}[t] 
	\includegraphics[width=0.9\columnwidth]{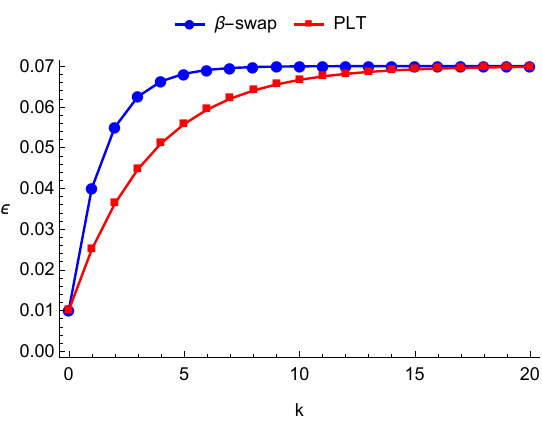}
	\caption{\emph{Comparison of 3 qubit protocols.} Here we show how the polarisation increases with number of iterations when the ETO is taken to be either a $\beta$-swap or a full thermalisation. One sees that the new protocol reaches the asymptotic polarisation value more rapidly than the original.}
	\label{fig:arbitraryn}
\end{figure}

Initially all $n$ qubits start in the thermal state. The steps are the following
\begin{enumerate}
	
	\item Apply the $n-1$ algorithm to all but the target qubit.
	\item Apply a Pauli-$X$ gate to each of these $n-1$ qubits.
	\item Apply an ETO to the lowest and maximally excited state of the $n$ qubits. That is, $\ket{0}_{S_0}.....\ket{0}_{S_{n-1}}$ and $\ket{1}_{S_0}.....\ket{1}_{S_{n-1}}$.
	\item Reset all the qubits but the target one in the thermal state.
	\item Repeat until asymptotic value is reached.
\end{enumerate}

It is shown in Ref.~\cite{rodriguez2017heat} that if the ETO is chosen to be a full thermalisation, then for general $n$ the asymptotic polarisation is given by
\begin{equation} \label{eq:asympn}
\epsilon^n_{\infty}=\tanh{\left[\left(2^{n-1}-\frac{1}{2}\right)\beta \Delta\right]}.
\end{equation}

What happens if the ETO is instead chosen to be a $\beta$-swap? It can be shown that in general the asymptotic value of Eq.~\eqref{eq:asympn} stays the same, but the convergence to it is still enhanced. We give a numerical example of this for the case of $n=3$ in Fig. \ref{fig:arbitraryn}. However, for small initial polarisation (and in contrast to the 2-qubit case), one run of the $n$ qubit algorithm utilizing a $\beta$-swap is not sufficient to achieve the asymptotic polarisation.

\section{Discussion and conclusions}

In this paper we have introduced the thermodynamic resource theory of elementary thermal operations (ETOs) and shown that for certain parameter regimes that they can be well reproduced by a simple Jaynes-Cummings interaction. While we proved that they are not as powerful as the much studied resource theory of thermal operations (TOs), both in terms of population dynamics and coherence manipulations, this link to the Jaynes-Cummings model makes ETOs more experimentally palatable than the complete set of TO. In doing so, we have show that natural analogues of powerful theorems from the theory of doubly stochastic matrices do not apply to finite temperature thermodynamics; physically this reflects a dependence of the thermodynamic constraints on the specific experimental restrictions at hand, in contrast to the large degree of universality observed in the infinite temperature limit. To overcome this we provide necessary and sufficient conditions for a given population dynamics to be achievable under ETO, but also some general tools to study more general restricted thermodynamic theories, including the resource theory of the Jaynes-Cummings model.

This set of results shows how natural limitations in the control assumed in a thermodynamic resource theory translate into stricter constraints than the standard thermodynamic single-shot quantities. For example, we have proven that for any $\v{p}$ for which the maximum extractable work under thermal operations is positive one has
	\begin{equation*}
	\begin{small}
	W_{\textrm{distil}}^{\textrm{JC}}(\v{p}) \leq 	W_{\textrm{distil}}^{\textrm{ETO}}(\v{p}) < 	W_{\textrm{distil}}^{\textrm{TO}}(\v{p}) = F_0(\v{p}) + \frac{1}{\beta} \log Z_S.
	\end{small}
	\end{equation*}	\noindent Hence, the \emph{min-free energy} $F_0(\v{p}) = - \beta^{-1} \log \sum_{i: p_i>0} e^{-\beta E_i}$ is not the thermodynamically relevant quantity in the theory of ETOs or the Jaynes-Cummings model. Similarly, we have seen that there are incoherent states that cannot be created under ETOs from a thermal state by adding work. Hence, for some $\v{p}$, $W^{\textrm{ETO}}_{\textrm{form}}(\v{p}) = +\infty$, even though \mbox{$W^{\textrm{ETO}}_{\textrm{form}}(\v{p}) = F_\infty(\v{p})+kT \log Z_S <+\infty$} for every $\v{p}$; this shows that the \emph{max-free energy} $F_\infty(\v{p}) = \beta^{-1} \log \max_i p_i e^{\beta E_i}$ is also not the thermodynamically relevant quantity in the restricted scenario.

It is worth comparing our framework with other approaches to providing experimentally feasible realisations of thermal operations. In Ref.~\cite{perry2015sufficient} it was shown that any population dynamics achievable using TO can also be produced by a protocol which first appends a single thermal qubit to the system and then applies sequences of partial levels thermalisations (PLTs) while slowly adjusting the energy levels of the combined system-qubit Hamiltonian (in a way judiciously chosen so as to not cost any work) before discarding the qubit in its original thermal state. While such a set of operations is more powerful than that considered here, in addition to the ability to selectively couple any pair of energy levels in the system to the environment (as is also the case for ETOs) it also requires the ability to precisely manipulate energy levels and thus adds an extra level of complexity to their implementation. This advantage suggests that NMR experiments may provide a platform to implement ETOs.

Since our necessary and sufficient conditions can be directly applied also to the Jaynes-Cummings model, even in parameter regimes where this is not well reproduced by ETOs, we can readily answer physically motivated questions. For example, in a cooling scenario one can ask: what is the maximum ground state population achievable by applying sequences and convex combinations of 2-level Jaynes-Cummings interactions in RWA, when the bath has temperature $T_B$? In Fig.~\ref{fig:JCmiliK} the answer corresponds to the ground state population of the extremal point closest to $(1,0,0)$. In the example of Fig.~\ref{fig:JCmiliK} the two qutrit transitions are \mbox{$\omega_{10} = 6.2$} GHz, \mbox{$\omega_{21} = 5.5$} GHz and the bath has temperature $T_B=0.1K$; if the initial state is \mbox{$\v{p}=(0.1,0.3,0.6)$}, we can bound the maximum final ground state population attainable through arbitrary Jaynes-Cummings interactions $p^{\rm max}_{\rm g}$ by $0.7607 <  p^{\rm max}_{\rm g} < 0.7757$. Note that this is much higher than the population that one would get by a trivial strategy consisting of simply thermalising the system with the bath at temperature $T_B=0.1K$, which gives $p^{\rm th}_{\rm g}\approx0.4921$. A generic sequence of ETOs can get to $p^{\rm max}_{\rm g} = 0.8061$, while a general thermal operation achieves $p^{\rm max}_{\rm g} = 0.895$, illustrating that the cooling power of the different sets is distinct, in accordance to the lack of universality discussed in Sec.~\ref{sec:lack}. Another crucial aspect is that, unlike other resource theory results, we know exactly the sequence of ETOs or Jaynes-Cummings interactions needed to get to the final state, so that we can devise explicit protocols.

Finally, there remain many open theoretical questions regarding ETOs. Firstly, it would be interesting to investigate whether a simple (and similarly experimentally motivated) operation can be added to them which allows the complete power of TOs to be realised. One could also investigate more general work extraction paradigms than the deterministic one consider here in Section~\ref{ssec:work}, such as those including a \emph{weight system} \cite{skrzypczyk2014work, alhambra2016fluctuating2}.

Thermodynamic resource theories have insofar been looking at fundamental constraints, but it is also necessary to close the gap with experimental implementations. We hope our contribution moves in this direction. While we have connected theory and physical models and gave tools to compute what protocols are possible, it remains to be seen how these results may help us devise efficient thermodynamical protocols for the quantum and nano scale. The improvements observed in the HBAC protocol gives some preliminary evidence in this direction.

\begin{acknowledgments}
We would like to thank D. Jennings, A.~Milne, A.~Levy, T.~Rudolph and  R.~Uzdin for comments on a previous draft and M.~Christandl, R.~Kosloff, N.~Yunger Halpern, M.~Horodecki, S.~F.~Huelga, P.~Mazurek, M.~Plenio, P.~Salamon and A.~Smith for useful discussions. ML is particularly grateful to M.~M\"uller for pointing out a mistake in a previous draft, which led to many developments, and to P.~Mazurek for interesting discussions about generalisations of Theorem~\ref{thm:extremal} to $k$-level operations. The result that ETO $\subset$ TO (which follows from Corollary \ref{corol:counter}) has been obtained independently by different methods by P. Mazurek and M. Horodecki in \cite{mazurek2017preparation}.  
ML acknowledges financial support from the Spanish MINECO (Severo Ochoa SEV-2015-0522 and project QIBEQI FIS2016-80773-P), Fundacio Cellex, and Generalitat de Catalunya (CERCA Programme and SGR 875).
CP acknowledges financial support from the European Research Council (ERC Grant Agreement no 337603), the Danish Council for Independent Research (Sapere Aude) and VILLUM FONDEN via the QMATH Centre of Excellence (Grant No. 10059).
AMA acknowledges support from the FQXi.
This work was supported by EPSRC and in part by COST Action MP1209.
\end{acknowledgments}
	
	\bibliographystyle{unsrtnat}
	
	\bibliography{Bibliography_thermodynamics}
	
\appendix

\section{Energy-preservation and complete passivity}
\label{appendix:energyconservation}
In this appendix, we briefly analyse why the condition \mbox{$[U,H_S + H_B] = 0$} is necessary for the (state-independent) set of allowed operations in Eq.~\eqref{eq:thermaloperations} to require no work to be performed. To do so, we will use the notions of passivity and complete passivity \cite{pusz1978passive, lenard1978thermodynamical}.

It is useful to rewrite the restriction on $U$ in an equivalent way. Let \mbox{$\gamma_\beta(H):=\frac{e^{-\beta H}}{\tr{}{e^{-\beta H}}}$}. Then \mbox{$[U,H_S + H_B] = 0$} is equivalent to
 	\be
  	U \left(\gamma_\beta (H_S) \otimes \gamma_\beta (H_B) \right) U^\dag = \gamma_\beta (H_S) \otimes \gamma_\beta(H_B).
 	\ee
 	Note that in the present manuscript we are assuming that the Hamiltonian of system and bath are the same at the end of the protocol as they are at the beginning (for generalisations, see Ref.~ \cite{brandao2011resource, horodecki2013fundamental}). If $U$ does not satisfy the previous equation, then it would be possible to transform the completely passive state $\gamma_\beta (H_S) \otimes \gamma_\beta (H_B)$ into one that is not. By applying such $U$ on several copies of \mbox{$\gamma_\beta (H_S) \otimes \gamma_\beta (H_B)$}, we would then get a state $\sigma$ that is not passive, i.e. a state from which it is possible to extract work. Specifically, there exists a time-dependent Hamiltonian $H(t)$ and \mbox{$\tau>0$} such that
 	\be
 	W = \int_{0}^\tau \tr{}{\sigma(t) \frac{dH(t)}{dt}}dt <0,
 	\ee 
 	where $\sigma(0) = \sigma$, $H(0) = H(\tau) = 0$ and $\sigma(t)$ is given by the Schr\"odinger equation for $H_S + H_B + H(t)$ \cite{lenard1978thermodynamical}. This tells us that an experimenter turning on the interaction $H(t)$ would be able to extract work from $\sigma$. Since no work could be extracted in this way from any number of copies of \mbox{$\gamma_\beta (H_S) \otimes \gamma_\beta (H_B)$}, by consistency the unitary $U$ must cost some work to perform, in contradiction to our assumption.

\section{Proof of Lemma~\ref{lem:eto}}
\label{appendix:elementary}

\eto*

\begin{proof}
	Let $\ket{E_0}$, $\ket{E_1}$ be two eigenstates of $H_S$ with energies $\hbar \omega_0$ and $\hbar \omega_1$, with $\omega_1 > \omega_0$. Introduce a single-mode thermal bath with Hamiltonian \mbox{$H_B = \sum_{n=0}^\infty n \hbar \omega \ketbra{n}{n}$}, with $ \omega = \omega_1- \omega_0$. Let the bath be in the state $\gamma_\beta(H_B)$, with $\gamma_\beta(H_B) = e^{-\beta H_B}/Z_B$, where $Z_B = (1-e^{-\beta \hbar \omega})^{-1}$. Take now the energy-preserving unitary (appearing in~\cite{aberg2013truly})
	\small{
		\begin{equation}
				\nonumber
		U = \ketbra{E_0}{E_0} \otimes \ketbra{0}{0} + \sum_{n=1}^\infty \sum_{k,k'=0}^1 v_{k k'} \ketbra{E_k}{E_{k'}} \otimes \ketbra{n-k}{n-k'},
		\end{equation}
	}where $V$ (whose elements are $v_{kk'}$) is a $2$ by $2$ unitary. Hence, by Eq.~\eqref{eq:thermaloperations}, the transformation
	\be
	\nonumber
	\mathcal{E}(\rho) = \tr{B}{U(\rho \otimes \gamma_B)U^\dagger},
	\ee 
	is an ETO. One can readily compute the induced population dynamics $G$ (defined as in Eq.~\eqref{eq:quantummapstostochasticmatrix})
	\be
	\nonumber
	G_{1|0} = e^{-\beta \hbar \omega}|v_{10}|^2, \quad G_{0|1} = |v_{01}|^2,		
	\ee
	and $G_{0|0} = 1- G_{1|0}$, $G_{1|1} = 1- G_{0|1}$. Since $V$ is unitary $|v_{10}|^2 = |v_{01}|^2$, thus $G_{1|0} = e^{-\beta \hbar \omega} G_{0|1}$. Given $G$ satisfying properties \ref{property1} and \ref{property2}, we can take $V$ of the form
	\be
	\nonumber
	V = \left[ \begin{matrix}
		\cos(x) & -i \sin (x) \\
		-i \sin(x) & \cos (x)
	\end{matrix} \right].
	\ee
	and choose $x$ such that $\sin^2(x) = G_{0|1}$.
\end{proof}

\section{\texorpdfstring{Upper bound on $G_{{\rm max}}(\bar{\beta})$}{Upper bound on G max}}
\label{appendix:bound}

	Consider the bound on $G_{0|1}(s)$ constructed following these steps:
	\begin{enumerate}
		\item Split the series in Eq.~\eqref{eq:JCtransitions} into a finite sum $F$ up to $m$, plus a residue $R$: $G_{0|1}(s) = F + R$, with \mbox{$F = \sum_{n=1}^m \sin^2(s \sqrt{n}) \frac{e^{-\beta \hbar \omega (n-1)}}{Z_B}$}.		
		In the residue, bound each factor $\sin^2(s\sqrt{n}) $ with $1$. Then one has $R= e^{-m \beta \omega}$.
		\item In $F$, bound every oscillating term with irrational frequency as  $\sin^2(s\sqrt{n}) \leq 1$.
	\end{enumerate}
	This provides a bound on $G_{0|1}(s)$ with a periodic function of $s$, which can then be simply analysed. For $m=4$, the function has one global maximum at $s= \pi/2$ when $\bar{\beta} \geq \log (4)/3$. On the other hand, when $\bar{\beta} < \log(4)/3$, it has two global maxima at $s= \pm \arccos(-e^{3\bar{\beta}}/4)/2$. The values achieved at the maxima give the bounds provided in the main text.

	\section{Collision models}
	\label{appendix:collision}
		Let the relevant two-level subsystem, with energies $\hbar \omega_0$, $\hbar \omega_1$, be described by the (possibly unnormalised) distribution \mbox{$\v{p}(0):=(p_0(0), p_1(0))$}. Without loss of generality, take \mbox{$\omega_0 = 0$} and \mbox{$\omega_1 = \omega$}. Let $\v{p}(n):=G^{n}\v{p}(0)$, where $G$ is the 2-level Gibbs-stochastic matrix induced by an individual collision. One can easily compute \mbox{$p_0{(1)} = p_0{(0)} (1-\lambda Z) + \lambda N$}, where $Z = 1+ e^{-\beta \hbar \omega}$, $N= p_0(0) + p_1(0)$ and $\lambda:=G_{0|1}$. By recursion, 
		\be
		\nonumber
		p_0{(n)} = p_0{(0)}(1-\lambda Z)^n + \frac{N}{Z}\left(1-(1-\lambda Z)^n\right).
		\ee
		Then \mbox{$p_0(\infty):=\lim_{n \rightarrow \infty} p_0(n) = N/Z$}, independently of the initial distribution. Because the map is stochastic, 
		\be
		\nonumber
		\v{p}(\infty) = N \v{g}
		\ee
		where $\v{g} = (1/Z, 1-1/Z)$ is a thermal distribution within the two-level subsystem.

		Reasoning as in Ref.~\cite{scarani2002thermalizing}, we can define $n = t/t_{{\rm int}}$, with $t_{{\rm int}}$ the duration of a single interaction. We then take the limit of short interactions $t_{{\rm int}} \rightarrow 0$, $\lambda \rightarrow 0$, while keeping $Z \lambda/ t_{{\rm int}} \rightarrow \xi \geq 0 $ finite ($Z$ is a constant that we absorbed in the definition of $\xi$). One can check that $(1-\lambda Z)^n \rightarrow e^{-t/\xi}$ and hence
		\be
		\label{eq:continuous}
		\v{p}(t) = e^{-t/\xi} \v{p}(0) + N (1-e^{-t/\xi})\v{g}.
		\ee
		Here $\xi \geq 0$ is a free parameter. Hence the subsystem decays exponentially to a distribution proportional to the Gibbs distribution (whenever $\xi>0$). 
	
\section{The TO Cone}
\label{appendix:TO Cone}

In this appendix we describe how to construct $\mathcal{C}_{\rm TO}(\v{p})$ given a state $\v{p}$.

Given a state $\v{p}$ and Hamiltonian $H_S$, let \mbox{$c_{\v{p}}:\left[0,Z_S\right]\rightarrow\left[0,1\right]$} be the function defined by the thermo-majorisation curve of $\v{p}$ i.e. $c_{\v{p}}\left(x\right)$ is the height of the thermo-majorisation curve of $\v{p}$ at $x$.
\begin{lem}
	Given $\v{p}$, consider the following distributions $\v{p}^\pi\in\mathcal{C}_{\rm TO}(\v{p})$ constructed for each permutation $\pi\in S_d$. For $i\in\left\{0,\dots,d-1\right\}$:
	\begin{enumerate}
		\item Let $x_i^{\pi}=\sum_{j=0}^{i} e^{-\beta E_{\pi^{-1}\left(j\right)}}$ and $y_i^{\pi}=c_{\v{p}}\left(x_i^{\pi}\right)$.
		\item Define $p^{\pi}_i:=y^\pi_{\pi(i)} - y^{\pi}_{\pi(i)-1}$, with $y_{-1}:=0$.
	\end{enumerate}
	Then all extremal points in $\mathcal{C}_{\rm TO}(\v{p})$ have the form $\v{p}^\pi$ for some $\pi$. This is in particular implies that $\mathcal{C}_{\rm TO}(\v{p})$ has at most $d!$ extremal points.
\end{lem}
Note that each $\pi\in S_d$ defines a different $\beta$-ordering on the energy levels of $H_S$. For each possible $\beta$-order, $\v{p}^\pi$ defines a thermo-majorization curve $c^{\pi}_{\v{p}}$ which coincides with the thermo-majorization curve of $\v{p}$ at the points $\left(x^{\pi}_i,y^{\pi}_i\right)$. In general, one has $c_{\v{p}}\left(x\right)\geq c_{\v{p}^{\pi}}\left(x\right)$ for every $x$ in $\left[0,Z_S\right]$ so $\v{p}^\pi\in\mathcal{C}_{\rm TO}(\v{p})$. Intuitively, $\v{p}^\pi$ is a distribution with $\beta$-ordering defined by $\pi$ that is ``just thermo-majorised'' by $\v{p}$. More precisely, there are no other distributions $\v{q}$ with the same $\beta$-order as $\v{p}^\pi$ whose thermo-majorisation curve lies under that of $\v{p}$ and above that of $\v{p}^\pi$.
\begin{proof}
We show that all extremal points of $\mathcal{C}_{\rm TO}\left(\v{p}\right)$ must have the above form. To see this, consider a state \mbox{$\v{q}\in\mathcal{C}_{\rm TO}\left(\v{p}\right)$} and suppose it were an extremal point. Let $\pi_q$ denote the $\beta$-order of $\v{q}$ and suppose that \mbox{$\v{q}\neq \v{p}^{\pi_q}$}. Now $\v{q}$ is thermo-majorised not only by $\v{p}$, but also by $\v{p}^{\pi_q}$. This follows from the fact that $\v{q}$ and $\v{p^{\pi_q}}$ have the same $\beta$-ordering and, furthermore, \mbox{$c_{\v{q}}(x^{\pi_q}_i) \leq c_{\v{p}}(x^{\pi_q}_i) = c_{\v{p}^{\pi_q}}(x^{\pi_q}_i)$.} Using Theorem 12 of Ref. \cite{perry2015sufficient}, we have that there exists a sequence of PLTs transforming $\v{p}^{\pi_q}$ into $\v{q}$. Using Eq.~\eqref{eq:etobetaswap}, we can then write $\v{q}=c_0 \v{p}^{\pi}+\sum_x c_x \beta^x  \v{p}^{\pi_q}$ where $\beta^x$ denotes a non-trivial sequence of $\beta$-swaps and $c_0>0$. As $\v{q}\neq \v{p}^{\pi_q}$, at least one other $c_x$ is non-zero. This contradicts $\v{q}$ being extremal.
\end{proof}
With this lemma in place, $\mathcal{C}_{\rm TO}(\v{p})$ can be found by constructing the states $\v{p}^\pi$ for each $\pi\in S_d$ and taking the convex hull.

		\section{Infinite work of formation: a counterexample not based on rank}
		\label{appendix:extracounterexample}
Here we show that the phenomena of infinite work of formation under ETO is not limited to the formation of states without full rank.
		
First note that as the initial state in formation processes is $\v{g}\otimes \ketbra{0}{0}$, without loss of generality we can begin by applying $\beta$-swaps between elements of $A_0$ and $A_w$ according to the injection $\Phi$ (here $\Phi$ is actually a bijection as $\v{g}$ has full support). To see this, note that applying $\beta$-swaps between elements of $A_0$ at any stage of the protocol will either do nothing (if both elements are occupied) or leave us unable to move all population out of $A_0$ (if one of the elements is unoccupied) as we will no longer be able to find a suitable injective map between $A_0$ and $A_w$. Similarly, applying $\beta$-swaps between elements of $A_w$ can either be postponed until we have applied all $\beta$-swaps associated with $\Phi$ (if both elements are occupied) or leave us unable to move all population out of $A_0$ (if one of the elements is unoccupied). Applying $\beta$-swaps according to $\Phi$ to $\v{g}\otimes\ketbra{0}{0}$ creates a state of the form $\v{g^{\pi}}\otimes\ketbra{1}{1}$ where $\v{g^{\pi}}$ is a permutation of $\v{g}$. Which permutations can be achieved is determined by the value of $w$. From this we can conclude that any $\v{p}$ such that $\v{g^{\pi}}\notetho\v{p}$ for any choice of $\pi$ we have  $W^{\textrm{ETO}}_{\textrm{form}}\left(\v{p}\right)=-\infty$. Examples of this include any state such that $p_{i}e^{\beta E_{i}}>\frac{e^{\beta\left(E_{d-1}-E_0\right)}}{Z_S}$, for some $i$.

\end{document}